\documentstyle[12pt,preprint]{aastex}  

\begin{document}

                           % The preamble begins here.
\title{An Infrared High Proper Motion Survey Using 2MASS and SDSS: Discovery of M, L and T Dwarfs}  
\author{Scott S. Sheppard}    
\affil{Department of  Terrestrial Magnetism, Carnegie Institution of Washington, \\
5241 Broad Branch Rd. NW, Washington, DC 20015 \\ sheppard@dtm.ciw.edu}

\and

\author{Michael C. Cushing}
\affil{Institute for Astronomy, University of Hawaii, \\
2680 Woodlawn Drive, Honolulu, HI 96822}
                 
\begin{abstract}  % Produces abstract

  A search of the Two Micron All Sky Survey and Sloan Digital Sky
  Survey reveals 36 previously unknown high proper motion objects with
  $J<17$.  Their red-optical colors indicate that 27 are M dwarfs, 8
  are early-type L dwarfs, and 1 is a late-type T dwarf.  The L dwarfs
  have $J-K_s$ colors near the extrema of known L dwarfs indicating
  that previous surveys for L dwarfs using color as a selection
  criterion may be biased.  Followup near-infrared spectroscopy of 6
  dwarfs confirm they are all late-type with spectral types ranging
  from M8 to T4. Spectroscopy also shows that some of the L dwarf
  spectra exhibit peculiar features similar to other peculiar "blue" L
  dwarfs which may indicate that these dwarfs have a relatively
  condensate free atmosphere or may be metal poor.  Photometric
  distance estimates indicate that 22 of the new M, L and T dwarfs lie
  within 100 pc of the Sun with the newly discovered T dwarf, 2MASS
  J10595185+3042059, located at $\sim$25 pc.  Based on the colors and
  proper motions of the newly identified objects, several appear to be
  good subdwarf candidates.  The proper motions of known ultracool
  dwarfs detected in our survey were also measured, including, for the
  first time, SDSS J085834.42$+$325627.6 (T1), SDSS
  J125011.65$+$392553.9 (T4) and 2MASS J15261405$+$2043414 (L7).

\end{abstract}


\section{Introduction}

Stars that exhibit high proper motions (HPMs: $\mu > 0\farcs4$
yr$^{-1}$) are likely to be within the solar neighborhood (less than
$\sim$100 pc).  Indeed most of our nearest stellar neighbors have been
identified through their proper motions using shallow optical surveys
such as the Digital Sky Survey (DSS) (Luyten 1979; Wroblewski and
Costa 2001; Ruiz et al. 2001; Oppenheimer et al. 2001; Lepine et
al. 2003; Teegarden et al. 2003; Hambly et al. 2004; Scholz et
al. 2004; Pokorny et al. 2004; Subasavage et al. 2005, Deacon et
al. 2005; Lodieu et al. 2005; Finch et al. 2007; Lepine 2008).
However, these surveys are not sensitive to the very red ($I-J > 4$),
low mass ultracool ($T_{\mathrm{eff}}<2400$ K) L and T dwarfs.

The L and T dwarfs are extremely faint in the optical and to date have
only efficiently been identified based on their colors at red optical
(0.6$-$1.0 $\mu$m) wavelengths in the Sloan Digital Sky Survey (SDSS:
e.g., Fan et al. 2000, Leggett et al. 2002, Chiu et al. 2006) and the
Canada France Brown Dwarf Survey (CFBDS: e.g., Delorme et al. 2008)
and at near-infrared (1$-$2.5 $\mu$m) wavelengths in the Two Micron
All Sky Survey (2MASS: e.g., Kirkpatrick et al. 1999, 2000; Burgasser
et al. 2002, 2003b; 2004a; Cruz et al. 2003; Tinney et al. 2005;
Kendall et al. 2007; Looper et al. 2007), the Deep Near Infrared
Southern Sky Survey (DENIS: Phan-Bao et al. 2008; Kendall et al. 2004)
and the UKIRT Infrared Deep Sky Survey (UKIDSS: e.g., Lawrence et
al. 2007; Chiu al. 2008, Pinfield et al. 2008).  Because these surveys
select ultracool dwarf candidates based on color alone, they may be
biased against objects with unusual colors.  Metchev et al. (2008)
have been able to partially relax this color constraint for low mass
objects by comparing the colors of objects in the 2MASS catalog to
colors of objects in the SDSS catalog.  Although such surveys have
proven very successful in identifying L and T dwarfs, it is likely,
based on the luminosity function of stellar and substellar objects,
that the census of the solar neighborhood within 25 pc remains roughly
20 to 60\% deficient (Henry et al. 1997; Henry et al. 2002; Reid et
al. 2004; Cruz et al. 2007; Reid et al. 2007).

A search for high proper motions stars at red optical and near
infrared wavelengths could 1) potentially identify brown dwarfs in the
solar neighborhood via their large proper motions, and 2) remove any
potential color bias inherent in most of the surveys conducted to
date.  Two surveys covering up to a few hundred square degrees have
been successful in identifying ultracool dwarfs via proper motion in
the near infrared (Artigau et al. 2006; Looper et al. 2008).  In this
paper, we present a large area survey ($\sim$ 8000 square degrees) to
search for faint, high proper motion stars in the near infrared using
the 2MASS and SDSS catalogs.

\section{Target Selection}

The 2MASS point source catalog (PSC: Skrutskie et al. 2006) between 0
and 66 degrees declination was compared to the SDSS catalog (York et
al. 2000; Pier et al. 2003) within this same range to search for HPM
objects.  Before comparing the 2MASS and SDSS databases, we performed
several cuts on the 2MASS PSC in order to limit the number of
candidate HPM objects: 1) The object must be detected in all three
filters ($J$, $H$ and $K_s$). 2) The $J$-band magnitude must be
brighter than 17th magnitude.  3) The distance between the source and
its nearest neighbor in the PSC must be greater than 5$\farcs$9 (a few
pixels) to prevent confusion (prox $> 5\farcs9$).  4) The
contamination and confusion flags must be ``0'' in all three bands to
prevent known artifacts or bright nearby objects from giving false
positives (cc\_flg = 000).  5) The extended source contamination flag
must be ``0'' to indicate the object does not fall within the
elliptical boundary of a known extended source (gal\_contam = 0).  6)
The minor planet flag must be ``0'' to identify known minor planets
(mp\_flg $=0$).  7) The object can not be identified with a known
object within 5$''$ from either the USNO-A2.0 or Tycho 2 catalogs (a
$= 0$).

The filtered 2MASS PSC was then compared to the SDSS catalog in order
to identify 2MASS point sources that did not have a SDSS source within
2$\farcs$35. This HPM survey covered area from the SDSS using data
releases one to five; DR1 ($\sim 2099$ square degrees; Abazajian et
al. 2003), DR2 ($\sim 1225$ square degrees; Abazajian et al. 2004),
DR3 ($\sim 1958$ square degrees; Abazajian et al. 2005), DR4 ($\sim
1388$ square degrees; Adelman-McCarthy et al. 2006), and DR5 ($\sim
1330$ square degrees; Adelman-McCarthy et al. 2007).  In all about
8000 square degrees of sky were covered by the SDSS between DR1 and
DR5.

\begin{figure}
\epsscale{0.4}
\centerline{\includegraphics[angle=90,width=4.5in]{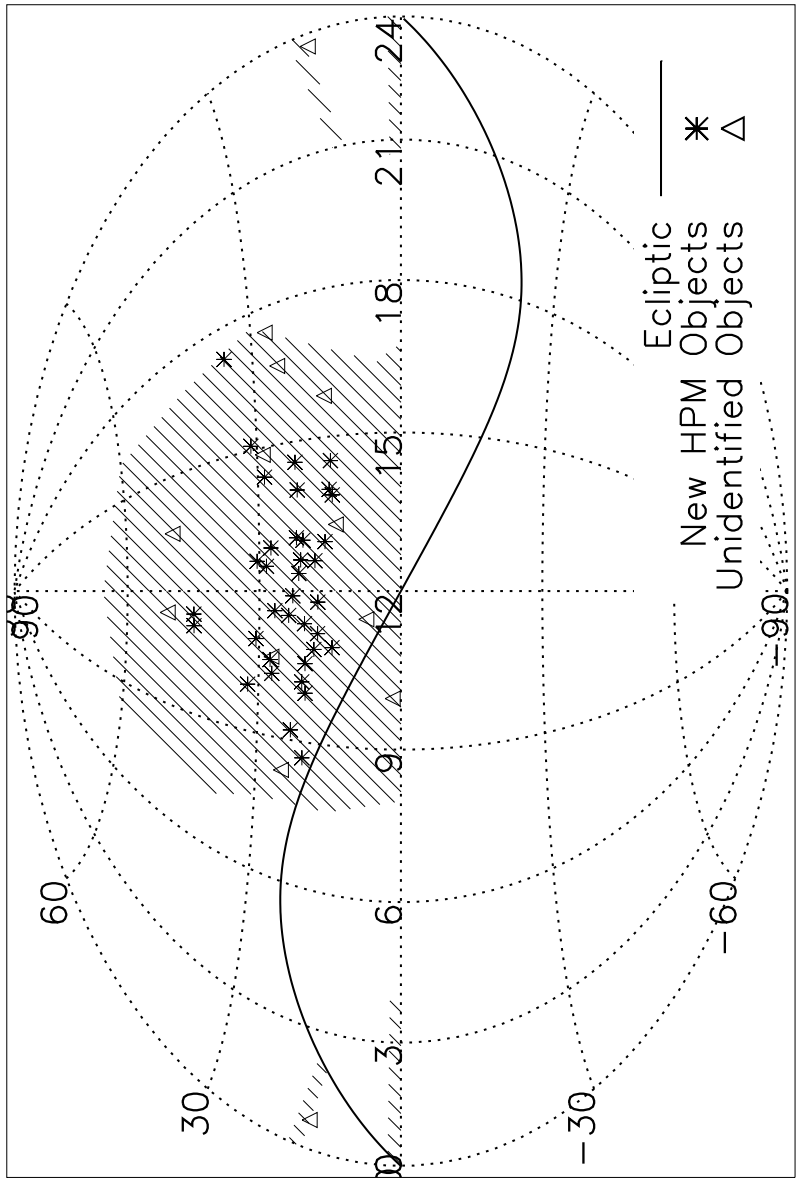}}
\caption{The location of newly identified high proper motion objects
  with declination in degrees on the vertical axis and right ascension
  in hours on the horizontal axis.  The shaded area is the 8000 square
  degrees where the 2MASS and SDSS data overlapped.  Most of the 36
  newly identified HPM objects are between about 15 and 30 degrees
  declination since this is where the SDSS Data Release 5 (DR5) was
  focused.  The DR5 has the longest time baseline between the 2MASS
  and SDSS images allowing for the more numerous slower proper motion
  objects to be identified. The 13 2MASS objects that do not have an
  identified SDSS source (open triangles) are likely artifacts in the
  2MASS catalog but could be unknown asteroids, extremely red objects,
  or extremely high proper motion objects.  Since the triangles are
  uncorrelated with the ecliptic (solid line) its unlikely they are
  unknown asteroids.}
\label{fig:2massmap} 
\end{figure}

The comparison between the 2MASS and SDSS surveys revealed 1826 objects
were cataloged by 2MASS but not cataloged by the SDSS.  1307 of these
objects were visually identified as stars in the SDSS images with no
detected proper motions between the SDSS and 2MASS data.  These stars
were not included in the SDSS catalog and thus resulted in false
positives.  The other 519 objects are candidate HPM objects that were
in the 2MASS PSC but had no point sources observed within 2$\farcs$35
in the SDSS data.  357 of the 519 were identified through SIMBAD as
known HPM stars and white dwarfs with motions between 0$\farcs$2 and
3$\farcs$9 yr$^{-1}$, 98 were identified as known asteroids that the
2MASS minor planet flag did not register using Lowell Observatory's
asteroid plot program (ASTPLOT; Granvik et al. 2003), 15 were
identified through SIMBAD and the Dwarf Archives\footnote{A current
list of all known L and T dwarfs is compiled at DwarfArcives.org} as
known L and T dwarfs (see Table 1).  13 objects (Table 2) could not be
correlated with any known SDSS objects or asteroids and thus they are
either: 1) Noise or artifacts in the 2MASS catalog, 2) unknown
asteroids, 3) extremely red objects that would not have been detected
in the $z'$, $i'$ or shorter wavelength regime of the SDSS, or 4)
extremely fast moving high proper motion objects of well over 10$''$
yr$^{-1}$, which is unlikely considering only one known star,
Barnard's Star, has a proper motion this large.  All 13 of the unknown
objects are very faint with $J$-band magnitudes well over the 2MASS
completeness limit of $J \sim 16$ mag.  The main belt asteroid
population at these faint magnitudes is probably close to complete
though Near Earth Objects (NEOs) are not.  As seen in
Figure~\ref{fig:2massmap}, there is no obvious clustering of
unidentified objects near the ecliptic as would be expected if most
were unknown asteroids.  The likely source for each of these
unidentified objects is given in Table 2.  In addition to comparing
the 2MASS and SDSS catalogs we used the Lowell asteroid ephemeris
service to determined that the hundreds of 2MASS sources that
Burgasser et al. (2003a, 2003b, 2004a) did not detect in their
follow-up imaging observations of candidate T dwarfs were indeed
asteroids with a few objects being obvious 2MASS artifacts.

The remaining 36 objects in our sample were found to have proper
motions between about 0$\farcs$2 and 1$\farcs$0 yr$^{-1}$ but were not
in SIMBAD (within 60$''$ of the predicted position) or the Dwarf
Archives as of 2008 May 1.  These 36 objects are thus identified as
new HPM objects.  Their Proper Motions (PM), Position Angles (PA), and
2MASS $J$,$H$,$K_s$ and SDSS $r'$, $i'$ and $z'$ photometry are given
in Table 3.  We discuss these 36 HPM objects in more detail in the
following sections.

\section{Analysis}
\subsection{Survey Results}

The proper motions of the 36 newly discovered HPM objects are all less
than 1$''$ yr$^{-1}$ although objects moving ten times faster would
likely have been detected or at least identified as having no SDSS
counterpart.  The fastest moving non Solar System object detected in
this survey was a white dwarf (GJ 518) with a motion of 3$\farcs$9
yr$^{-1}$.  Several more known objects, including L and T dwarfs (Table
1), moving well over 1$''$ yr$^{-1}$ were also detected.

\begin{figure}
\epsscale{0.7}
\centerline{\includegraphics[angle=90,width=\textwidth]{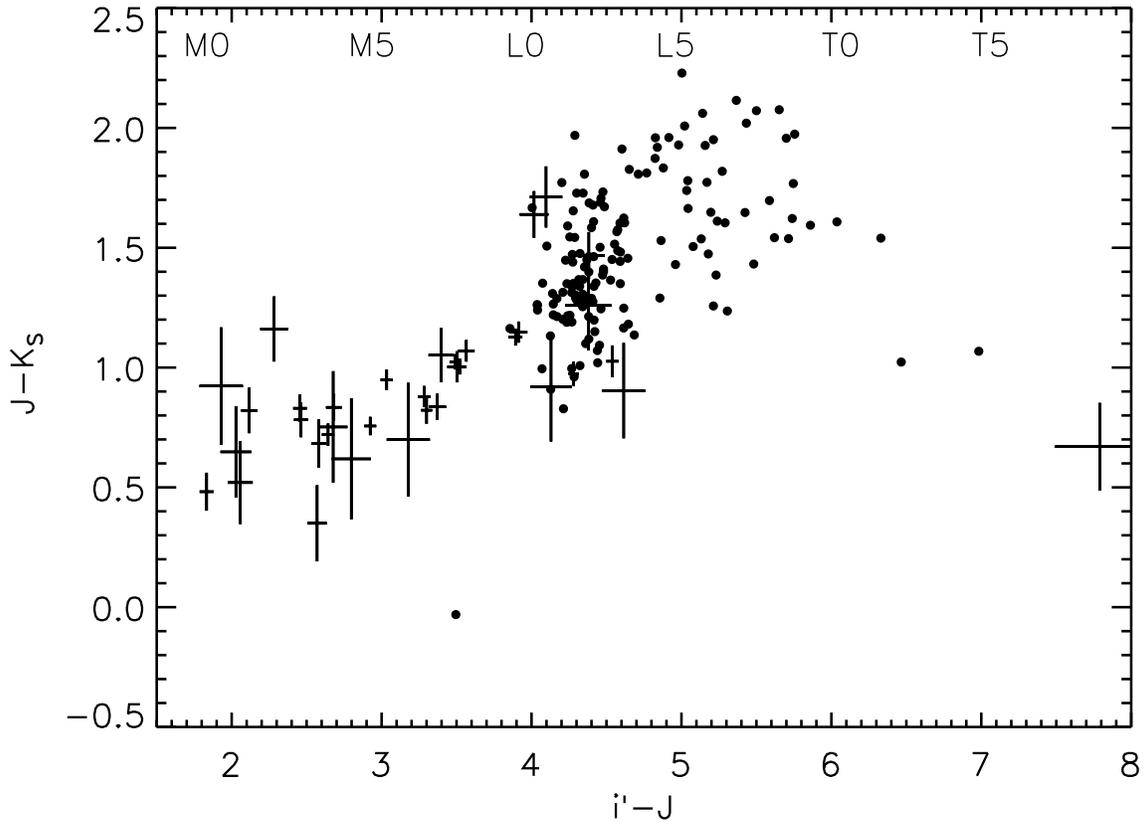}}
\caption{The $i'-J$ versus the $J-K_s$ colors of the 36 newly
  discovered high proper motion objects (plus signs).  Also plotted
  are the known ultracool dwarfs with spectral types L0 or later
  observed in the SDSS (filled circles) as of 2008 May 1.  All colors
  were obtained through the 2MASS and SDSS point source catalogs.
  Approximate spectral types based on the average $i'-J$ colors of M,
  L, and T dwarfs are given (see text).  It is apparent that most of
  the new candidate L dwarfs have $J-K_s$ colors that lie at the
  extrema of the known early-type L dwarfs suggesting a possible bias
  in the sample of known L dwarfs that were identified based on their
  colors alone.  The T dwarf candidate is at the right.  T dwarfs are
  extremely faint in the $i'$ band and thus very few have been
  observed in the SDSS.  The known brown dwarf in the lower center of
  this figure is the L4 subdwarf, 2MASS J16262034$+$3925190 (Burgasser
  et al. 2004b).}
\label{fig:ijjkcolors} 
\end{figure}

\begin{figure}
\epsscale{0.7}
\centerline{\includegraphics[angle=90,width=\textwidth]{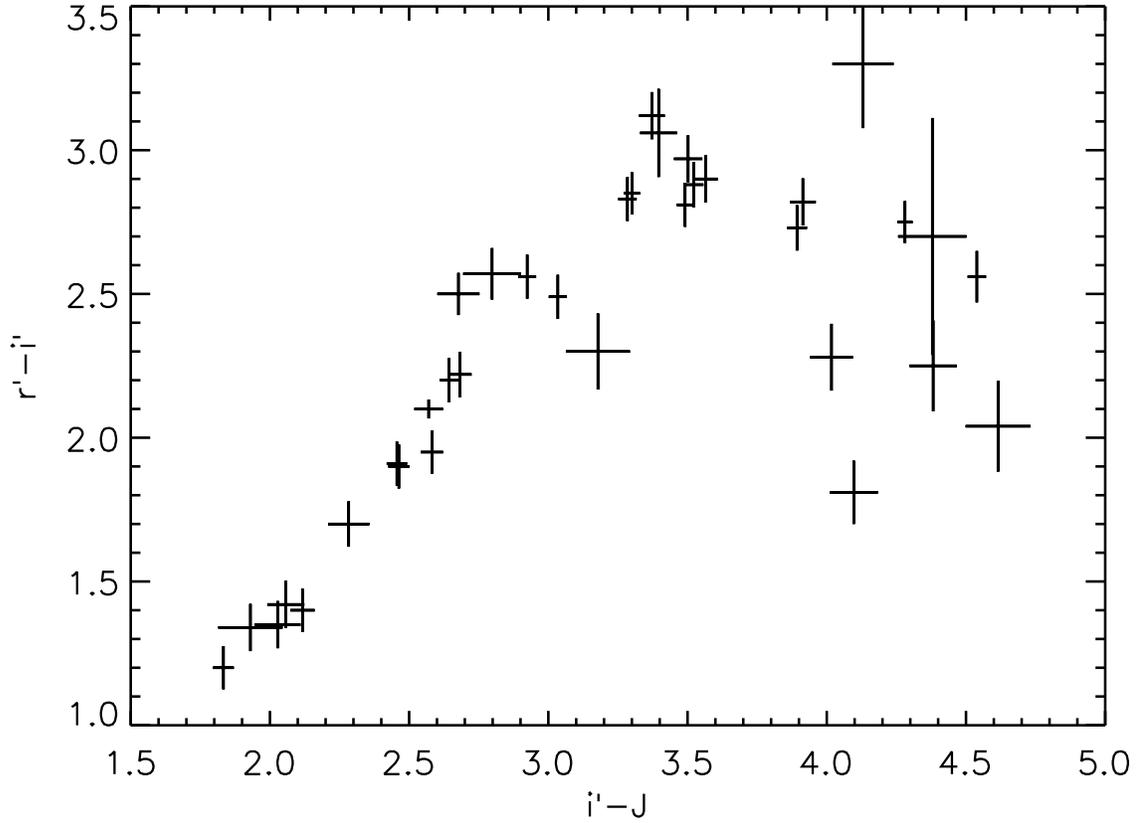}}
\caption{The $i'-J$ versus $r'-i'$ colors of the 36 newly discovered
  high proper motion objects.  There is an obvious turn over near
  $r'-i' \sim$3 and $i'-J \sim$ 4.  This turn over has been associated
  with Ti-bearing condensates that begin in late-type M dwarfs
  (Kirkpatrick et al. 2000; Dahn et al. 2002; Liebert and Gizis
  2006). The T dwarf candidate is not in this figure because its $r'$
  magnitude is unknown.}
\label{fig:ijricolors} 
\end{figure}

\begin{figure}
\epsscale{0.7}
\centerline{\includegraphics[angle=90,width=\textwidth]{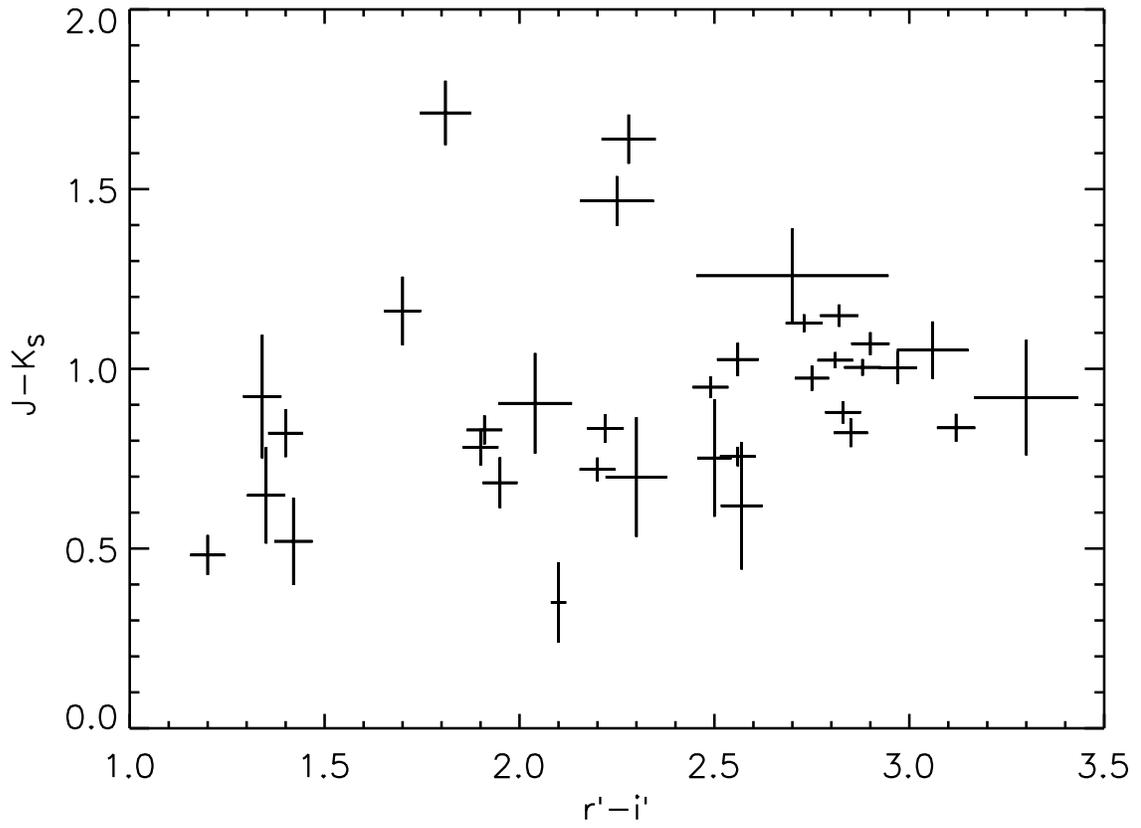}}
\caption{The $r'-i'$ versus $J-K_s$ colors of the 36 newly discovered
  high proper motion objects.  The T dwarf candidate is not in this
  figure because its $r'$ magnitude is unknown.}
\label{fig:rijkcolors} 
\end{figure}

All the newly discovered HPM objects are quite faint, with many being
fainter than 20th magnitude in $r'$ and thus not efficiently detected
in earlier optical surveys like the DSS.  The colors of the new HPM
objects are given in Table 4.  $J-K_s$ versus $i'-J$, $r'-i'$ versus
$i'-J$, and $J-K_s$ versus $r'-i'$ color-color diagrams are shown in
Figures~\ref{fig:ijjkcolors} to~\ref{fig:rijkcolors}.  The large
scatter in the $J-K_s$ colors of the L dwarfs
(Figure~\ref{fig:ijjkcolors}) has been ascribed to variations in the
properties of their condensate clouds (Tsuji et al.1996; Knapp et
al. 2004), surface gravities (Burrows et al. 2006) and possible
unresolved binaries (Liu and Leggett 2005).  This scatter has been
found to be even higher for the lower temperature T dwarfs (Dahn et
al. 2002; Harris et al. 2003).  Figure~\ref{fig:ijricolors} shows a
significant turn over in the object's $r'-i'$ colors resulting from
the formation of Ti-bearing condensates that begin in the late-type M
dwarfs (Kirkpatrick et al. 2000; Dahn et al. 2002; Liebert and Gizis
2006).

\begin{figure}
\epsscale{0.7}
\centerline{\includegraphics[angle=90,width=\textwidth]{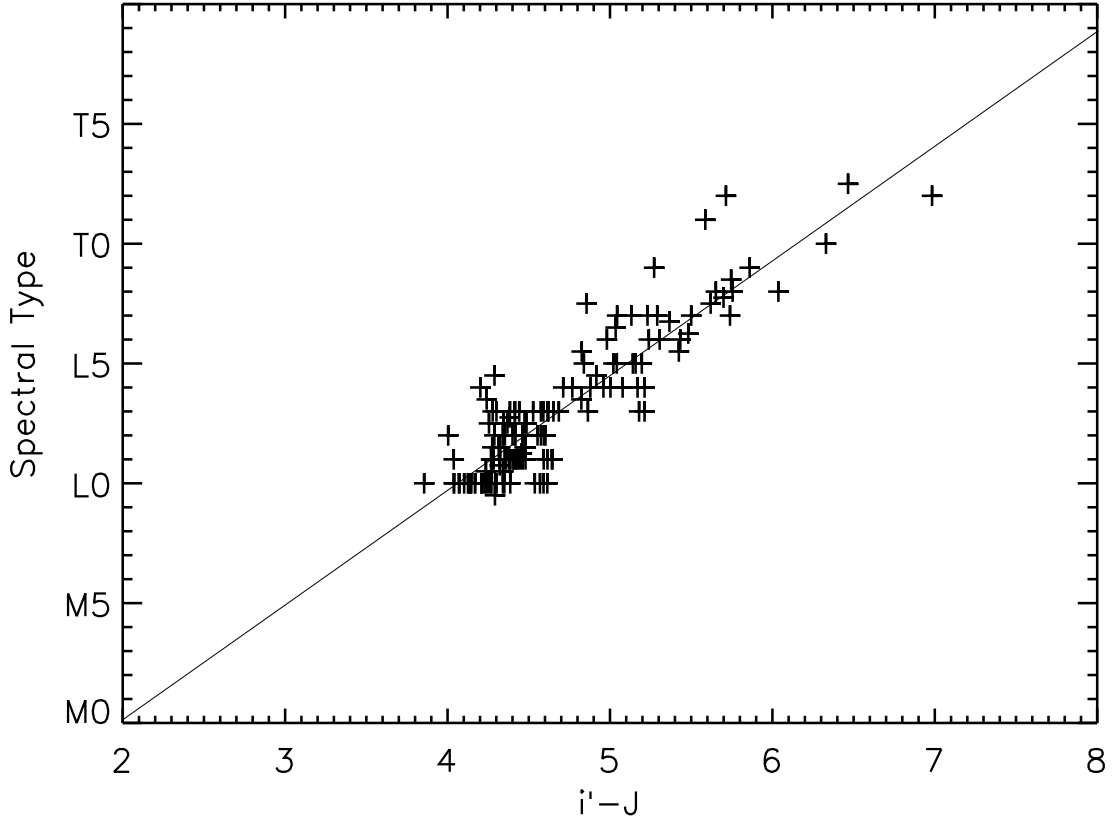}}
\caption{The $i'-J$ colors versus spectral type of known L and T
  dwarfs detected by the SDSS.  All ultracool dwarfs plotted had their
  photometry taken from the SDSS and 2MASS data bases exactly as the
  36 newly discovered objects found in this work to allow a direct
  comparison.  The solid line is the best fit to the data and is used
  to estimate the spectral types of the newly observed objects (Table
  4). The linear spectral type fit found for brown dwarfs using i'-J
  is: $\mbox{SpecType} = -9.4 + 4.8(\mbox{i'-J})$, where the Spectral
  Type is 05=M5, 10=L0, 15=L5, 20=T0, 25=T5 etc.  Because T dwarfs are
  extremely faint in the i' band few have been reliably observed in
  the SDSS.  As shown by Hawley et al. (2002) this linear fit is good
  down to the earliest M type dwarfs.  We extrapolate the spectral
  type based on the linearity of the fit for objects that have a
  $i'-J$ color less than 3.9 and greater than 7 magnitudes.}
\label{fig:typeIJ} 
\end{figure}

The rough spectral types for the new HPM objects were estimated using
their $i'-J$ colors (Table 4) and a linear fit to known ultracool
dwarfs with known spectral types as shown in the Dwarf Archives
(Figure~\ref{fig:typeIJ}).  We found that the spectral type of an
ultracool dwarf is shown to follow the relation
\begin{equation}
\mbox{SpecType} = -9.4 + 4.7(\mbox{$i'-J$})
\label{eq:spectype}
\end{equation}
\noindent
where SpecType is the spectral type of the object with 10=L0, 15=L5,
20=T0, 25=T5 etc.  The average spectral typing error using
Equation~\ref{eq:spectype} (i.e. the $i'-J$ information) on the sample
of objects shown in Figure~\ref{fig:typeIJ} is about $\pm 1$ sub
class.  Equation~\ref{eq:spectype} is nearly linear down to at least
the early M types ($i'-J\sim 2$; Hawley et al. 2002).  Thus even
though Equation~\ref{eq:spectype} was determined based on L and
early-type T dwarf data covering from $i-J\approx 3.9-7.0$, we can
extrapolate to bluer colors in order to spectral type M dwarfs.

\begin{figure}
\epsscale{0.7}
\centerline{\includegraphics[angle=90,width=\textwidth]{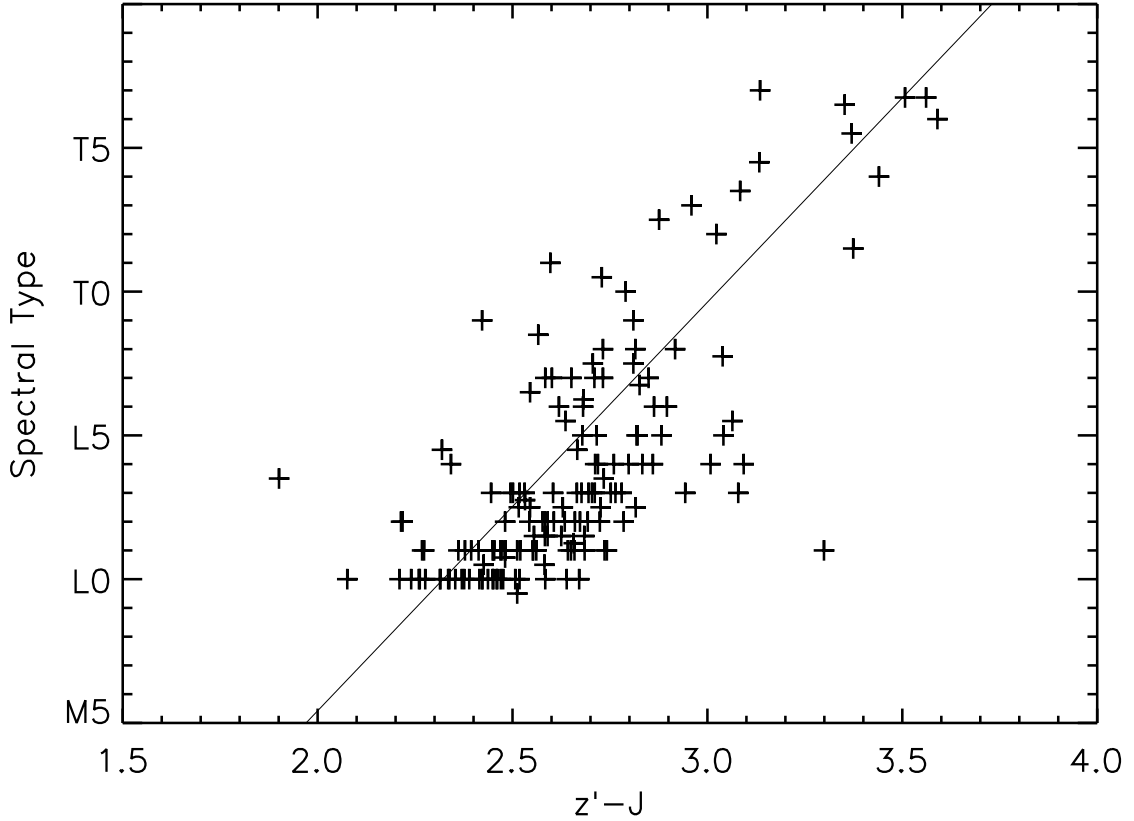}}
\caption{The $z'-J$ colors versus spectral type of known L and T
  dwarfs observed with the SDSS and 2MASS (see dwarfarchives.org). The
  solid line is the best fit to the data and is used to estimate the
  spectral type of the newly discovered T dwarf 2MASS
  J10595185$+$3042059 (Table 4).  Determining the spectral type
  through the $z'-J$ color technique is good for late-type T dwarfs
  because $i'$ band detections are rare in the SDSS.  The linear
  spectral type fit found for brown dwarfs using z'-J is:
  $\mbox{SpecType} = -23.0 + 14.2(\mbox{z'-J})$, where the Spectral
  Type is 05=M5, 10=L0, 15=L5, 20=T0, 25=T5 etc.}
\label{fig:typeZJ} 
\end{figure}

Since the extremely red ($i'-J \sim$7.8) object 2MASS
J1059$+$3042\footnote{Hereafter we abbreviate the 2MASS and SDSS
designations as hhmm$\pm$ddmm where the suffix is the sexagesimal
right ascension (hours and minutes) and declination (degrees and
arcminutes) at J2000.0 equinox.} was not detected in the SDSS $i'$
band we estimated its spectral type using its $z'-J$ color.  To do so
we determined a relation for spectral type versus $z'-J$ colors
(Figure~\ref{fig:typeZJ})
\begin{equation}
\mbox{SpecType} = -23.0 +14.2(\mbox{$z'-J$})
\label{eq:spectypezj}
\end{equation}
\noindent
where SpecType is the same as above.  This technique is inferior to
the $i'-J$ colors technique used above since the z'-band is closer in
wavelength to the J-band but is useful for extremely red objects that
are difficult to observe in the $i'$-band.  The average spectral
typing error using Equation~\ref{eq:spectypezj} on the sample of
objects shown in Figure~\ref{fig:typeZJ} is about $\pm 3$ sub classes.
The spectral types found using Equation~\ref{eq:spectype} or
Equation~\ref{eq:spectypezj} for each of the 36 new HPM objects are
given in Table 4.

\begin{figure}
\epsscale{0.7}
\centerline{\includegraphics[angle=90,width=\textwidth]{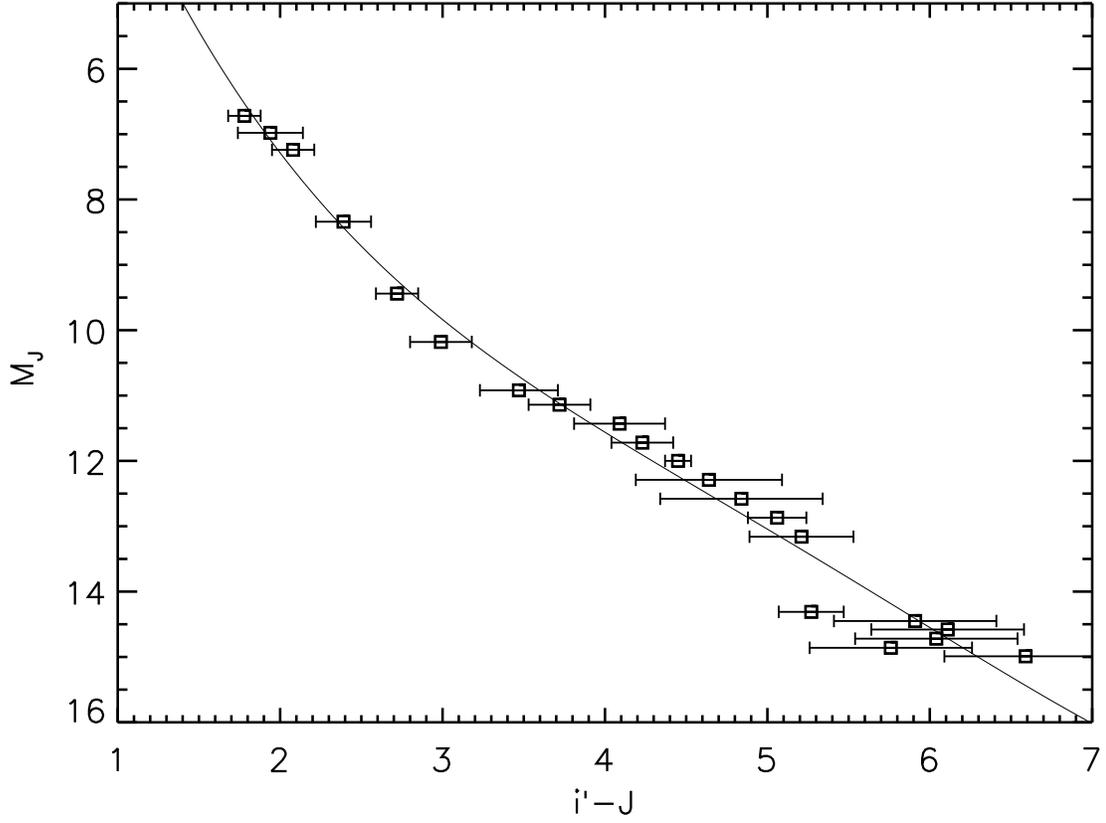}}
\caption{The fitted $i'-J$ versus absolute magnitude using data points
  from Hawley et al. (2002).  This fit is used to determine the
  absolute magnitude, $M_{J}$, of the newly identified high proper
  motion objects using their $i'-J$ magnitudes: $M_{J} = -9.66 +
  15.54(\mbox{$i'-J$}) - 4.81(\mbox{$i'-J$})^{2} +
  0.72(\mbox{$i'-J$})^{3} - 0.04(\mbox{$i'-J$})^{4}$ }
\label{fig:hpabsIJ} 
\end{figure}

The approximate absolute magnitude, $M_{J}$, of the new HPM objects were
estimated by using their $i'-J$ colors and data from Hawley et
al. (2002) (see Figure~\ref{fig:hpabsIJ})
\begin{equation}
M_{J} = -9.66 + 15.54(\mbox{$i'-J$}) - 4.81(\mbox{$i'-J$})^{2} + 0.72(\mbox{$i'-J$})^{3} - 0.04(\mbox{$i'-J$})^{4}
\label{eq:absolutemag}
\end{equation}
\noindent
and are given in Table 4.  Our absolute magnitude calculations are
similar to ones used by Cruz et al. (2003) and Phan-Bao et al. (2007)
but our analysis allows us to directly relate the $i'-J$ colors to
$M_{J}$ for these newly observed ultracool dwarfs rather than using
their unknown spectral types.  The absolute magnitudes were used to
determined the approximate distances to the objects through,
$d=10^{(J-M_{J}+5)/5}$ pc.  These are very rough estimates for the new
HPM objects identified in this survey and assume the objects are
single.  The distance estimates have uncertainties at about the 25
percent level based on comparing the results with known objects in
Table 1.  All the objects later than M8 appear to be less than about
100 parsecs from the Earth (Table 4).  Several of the early M dwarfs
that appear to be at very large distances would have unrealistic
tangential velocities ($\gtrsim$ 500 km s$^{-1}$).  These objects are
good subdwarf candidates (see section 4) and thus their absolute
magnitudes and distances are likely unreliable since the distance
estimates assume the objects to be dwarfs and not subdwarfs.  The
closest object found in this survey is the T dwarf 2MASS J1059$+$3042
which lies at a distance of $\sim$25 pc if it is a single brown dwarf.

\subsection{Survey Consistency}

The 2MASS data were taken between 1997 and 2001 (Skrutskie et al. 2006)
while the SDSS data acquisition started in 2000 with DR1 and went
through DR5, which ended in 2005.  Thus the time baseline between DR1
and 2MASS is only a few years (1 to 3 years) while the time baseline for
DR5 and 2MASS is several years (up to 8 years).  Since most of the new
HPM objects are only moving a few tenths of arcseconds per year, they
can only be detected in our analysis with a long time baseline and as a
result, almost all the new HPM objects discovered in this work were
found in the DR5 data release (Figure~\ref{fig:2massmap}).

As a consistency check we identified 32 L and T dwarfs in the Dwarf
Archives with HPM ($>0\farcs$4 yr$^{-1}$) that our survey should have
detected.  We found 15 of these objects (Table 1).  We determined that
7 of the 17 objects were not seen in our survey because they had no
obvious motion between the 2MASS and SDSS images (the images were
taken too close in time) and thus would not be expected to be detected
in our survey.  One of the 17 objects was located close (about 5
arcminutes) to a nearby bright star ($J \sim 1$ mag) and thus was not
expected to be detected by our survey because of the large amount of
scattered light.  Four of the 17 were not clearly detected in the
$K_s$ band in the 2MASS data and thus were rejected by our search
algorithm.  Five of the 17 objects were closer than 2$\farcs$35 to
another source in the SDSS data and thus also would have been rejected
by our search algorithm.  In total we found 100\% (15/15) of the known
L and T dwarfs that our survey was expected to find.  As another
measure of the efficiency of our survey, we found about 60\% (15/25)
of the known L and T dwarfs within our survey that had obvious motion
between the 2MASS and SDSS data ($> 2''$).

\subsection{Followup Observations}

\subsubsection{du Pont}

We observed all 36 objects listed in Table 4 with a Tektronix 2048x2048
CCD ($0\farcs$259 per pixel) on the du Pont 2.5 meter telescope at Las
Campanas, Chile in 2007 Dec, 2008 Feb, and 2008 Mar.  These 200 sec
$I$-band images confirmed 35 of these objects as having high proper
motions.  The only object not seen in the $I$-band du Pont data was
2MASS J1059$+$3042.  We thus observed this object in 2008 May with
LDSS-3 on the 6.5 meter Magellan telescope in both the $i'$- and
$z'$-bands.  2MASS J1059$+$3042 was easily detected and confirmed as a
high proper motion source in two 100 sec $z'$-band exposures.  2MASS
J1059$+$3042 was only marginally detected in one 200 sec $i'$-band
exposure demonstrating the extreme redness ($i'-J \sim 7.8$) of this object
(Table 4).  In addition, we observed all the unidentified objects listed
in Table 2 with the du Pont telescope to search for faint $I$-band
counterparts that the SDSS would not have detected.  These 200 to 600
sec exposures found no obvious objects of interest near any of the
locations listed in Table 2.

\subsubsection{IRTF/SpeX}

Near-infrared spectra of six of the new HPM objects with estimated
spectral types later than M9 V were obtained with SpeX (Rayner et
al. 2003) mounted on the 3m NASA Infrared Telescope Facility on 2008
June 13 and 18 (UT).  A log of the observations, including the UT date,
total integration time, and A0 V standard star are given in Table 5.  We
used the 0$\farcs$5-wide slit with the low-resolution prism mode which
provides a resolving power of $R\equiv \lambda/\Delta \lambda \approx
150$.  A series of 120 sec exposures were obtained at two different
positions along the 15$''$-long slit to facility sky subtraction.  An A0
V standard star with a similar airmass ($\Delta \sec z$ $<$ 0.1) was
observed after each science target for telluric correction and flux
calibration purposes.  All observations were conducted at the
parallactic angle in order to minimize slit losses and spectral slope
variations due to differential atmospheric refraction.  Finally,
exposures of internal flat field and Ar arc lamps were obtained for
flat-fielding and wavelength-calibration purposes.

\begin{figure}
\epsscale{0.7}
\centerline{\includegraphics[angle=0,width=4in]{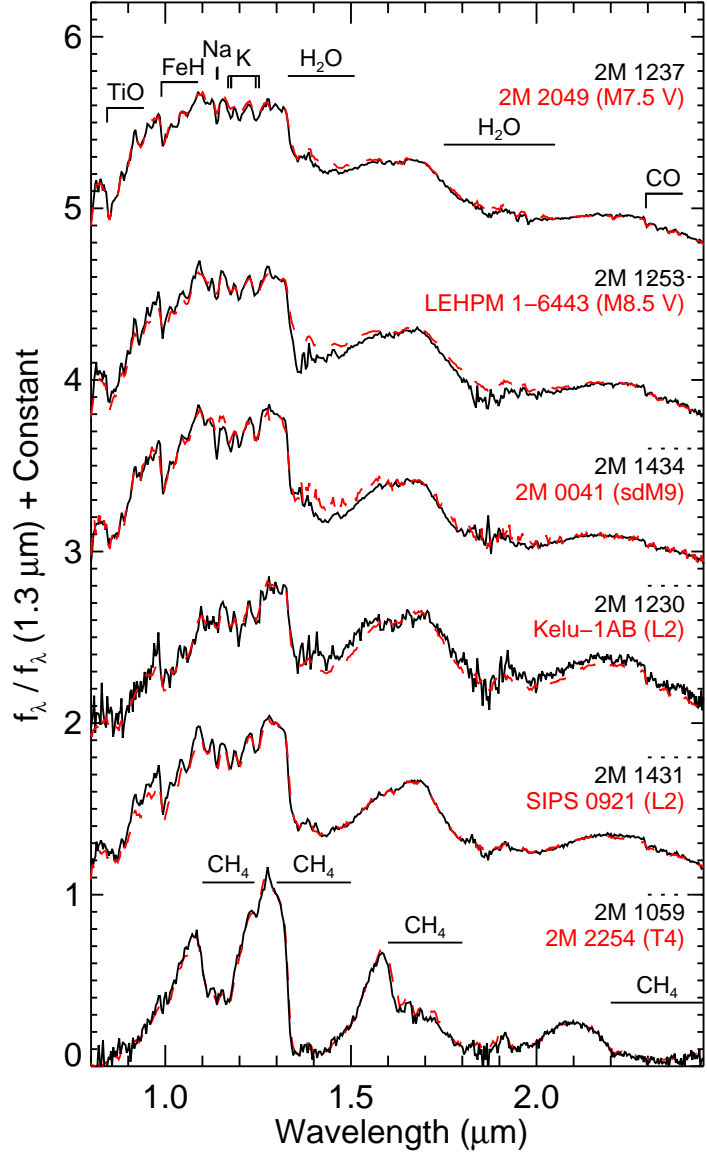}}
\caption{The 0.8$-$2.45 $\mu$m spectra (black) of 2MASS J1237$+$3028,
  2MASS J1253$+$2728, 2MASS J1434$+$2202, 2MASS J1230$+$2827, 2MASS
  J1431$+$1436, and 2MASS J1059$+$3042.  Overplotted (red) are the best
  fitting spectra in the SpeX Prism Spectral Library, 2MASS
  J2049$-$1944 (Burgasser et al. 2004a), LEHPM 1$-$6443 (Burgasser
  (2008b), 2MASS J0041$+$3547 (Burgasser et al. (2004a), Kelu-1AB
  (Burgasser et al. 2007b), SIPS 0921$+$2104 (Burgasser et al. 2007b)
  and 2MASS 2254$+$3123 (Burgasser et al. (2004a).  The spectra are
  normalized to unity at 1.3 $\mu$m and offset by constants (dotted
  lines).  Prominent atomic and molecular absorption features are
  indicated.}
\label{fig:tdwarfspec} 
\end{figure}

The data were reduced using Spextool, the data reduction package for
SpeX (Cushing et al. 2004).  The package performs nonlinearity
corrections, flat-fielding, optimal extraction of spectra, and
wavelength calibration.  The spectra were then corrected for telluric
absorption and flux calibrated using the observed A0 V standard star and
the technique described in Vacca et al. (2003).  The spectra of the 6
candidates are shown in Figure \ref{fig:tdwarfspec}.  The S/N of the
spectra range from 20 to 200 except for 2MASS J1230$+$2827 which
ranges from 10 to 50.  Five of the six dwarfs exhibit features typical
of late-type M and L dwarfs including broad H$_2$O absorption at
$\sim$1.4 and $\sim$1.8 $\mu$m, the Wing-Ford FeH band head at 0.99
$\mu$m, \ion{K}{1} and \ion{Na}{1} doublets from 1.15 to 1.3 $\mu$m, and
the CO overtone band heads at $\sim$2.29 $\mu$m.  The extremely red
object ($i'-J \sim$ 7.8), 2MASS J1059$+$3042, exhibits CH$_4$ absorption
bands at 1.2, 1.6, and 2.2 $\mu$m indicating that it is a T dwarf.

Spectral types for the 6 dwarfs were determined in two ways.  First,
each spectrum was compared to the $\sim$250 spectra of M, L, and T
dwarfs (obtained with the same instrument setup as our data) in the SpeX
Prism Spectral Library\footnote{http://www.browndwarfs.org/spexprism}.
The best fitting library spectrum for each dwarf was found by minimizing
the sum of the squared residuals between our spectrum and the library
spectra (after all spectra are normalized to unity over the 2.1 to 2.2
$\mu$m wavelength range) and are overplotted in Figure
\ref{fig:tdwarfspec}.  2MASS J1059$+$3042 is well matched by the T4
near-infrared standard 2MASS J2254188$+$312349 (Burgasser et al. 2006)
so we assign it a spectral type of T4.  The remaining 5 dwarfs are well
matched by their respective library spectra, but assigning them spectral
types by direct comparison is difficult since there are no spectral
standards for M and L dwarfs in the near-infrared; the spectral types
given in Figure \ref{fig:tdwarfspec} were derived from red optical data.
We have therefore also derived the spectral types of the 5 M and L
dwarfs using various spectral indices from the literature (Reid et
al. 2001, Testi et al. 2001, Burgasser et al. 2002, 2006, 2007b, 2008,
Allers et al. 2007) as well as the Geballe et al. (2002) indices that
are used to define near-infrared L dwarf subtypes.  The resulting
spectral types are given in Table \ref{tab:Indices} and a summary of the
spectral types derived by the various methods are given in Table
\ref{tab:SpTypes}.  With the exception of 2MASS J1059$+$3042, the
spectral types derived by the various methods are in reasonable
agreement.

\section{Discussion}

The original goal of this survey was to identify brown dwarfs that
were close to the Sun and to understand whether or not there is a
color bias in previous surveys that use color as a selection
criterion.  Although no brown dwarfs located very close to the Sun
were identified, our results suggest that a color bias does exist in
current surveys.  As shown in Figure ~\ref{fig:ijjkcolors}, six of the
eight L dwarfs identified in this survey have $J-K_s$ colors that lie
at the extrema of the distribution of L dwarfs colors.  Followup
spectroscopy shows that 2MASS J1431$+$1436 and 2MASS J1434$+$2202
appear to have anomalous spectral features given their (tentative)
spectral types.  Based on their $J-K_s$ colors alone, both dwarfs
would have a spectral type of earlier than M9 V (Kirkpatrick 2008)
indicating that these dwarfs are much bluer than typical L dwarfs.
Indeed Burgasser et al. (2004a, 2008a) have noted that the spectra of
2MASS J0041$-$3547 and SIPS J0921$-$2104, the best fitting library
spectra for 2MASS J1431$+$1436 and 2MASS J1434$+$2202 respectively,
appear anomalous with stronger FeH absorption, weaker CO absorption,
and a bluer near-infrared continuum than typical L dwarfs.  Such
spectral characteristics have been previously ascribed to subsolar
metallicities (e.g., Cruz et al. 2007), high surface gravities or
variations in the condensate clouds properties (e.g., Burgasser et
al. 2008a).  A sub solar metallicity may explain the peculiar nature
of 2MASS J1434$+$2202 given that the spectral type of its best fitting
library spectrum, 2MASS J0041$-$3547, is sdM9.

Most of the newly discovered objects are relatively distant (Table 4).
It is thus likely that most of these objects have large tangential
velocities in order for them to show such high proper motions at
relatively large distances.  Metal poor subdwarfs are usually
associated with large tangential velocities.  In order to determine if
any of the objects reported in this survey are subdwarfs we calculated
their Reduced Proper Motions, RPM.  The RPM of each object was
determined using the relation $RPM = r' + 5 + 5 \log \mu$ (Salim and
Gould 2002; Subasavage et al. 2005; Marshall 2008).  The RPM can be
thought of as a rough luminosity class (or absolute magnitude) using
only proper motion ($\mu$) and apparent magnitude ($r'$) by assuming
they are directly related.  This assumption is usually valid though
many factors make the scatter in the relationship large.  The RPM is a
simple technique to allow identification of object types that differ
from the normal main sequence.  This technique is useful when looking
for a crude way to identify white dwarfs and subdwarfs since they will
have bluer colors or fainter RPMs for a given color.

\begin{figure}
\epsscale{0.7}
\centerline{\includegraphics[angle=90,width=\textwidth]{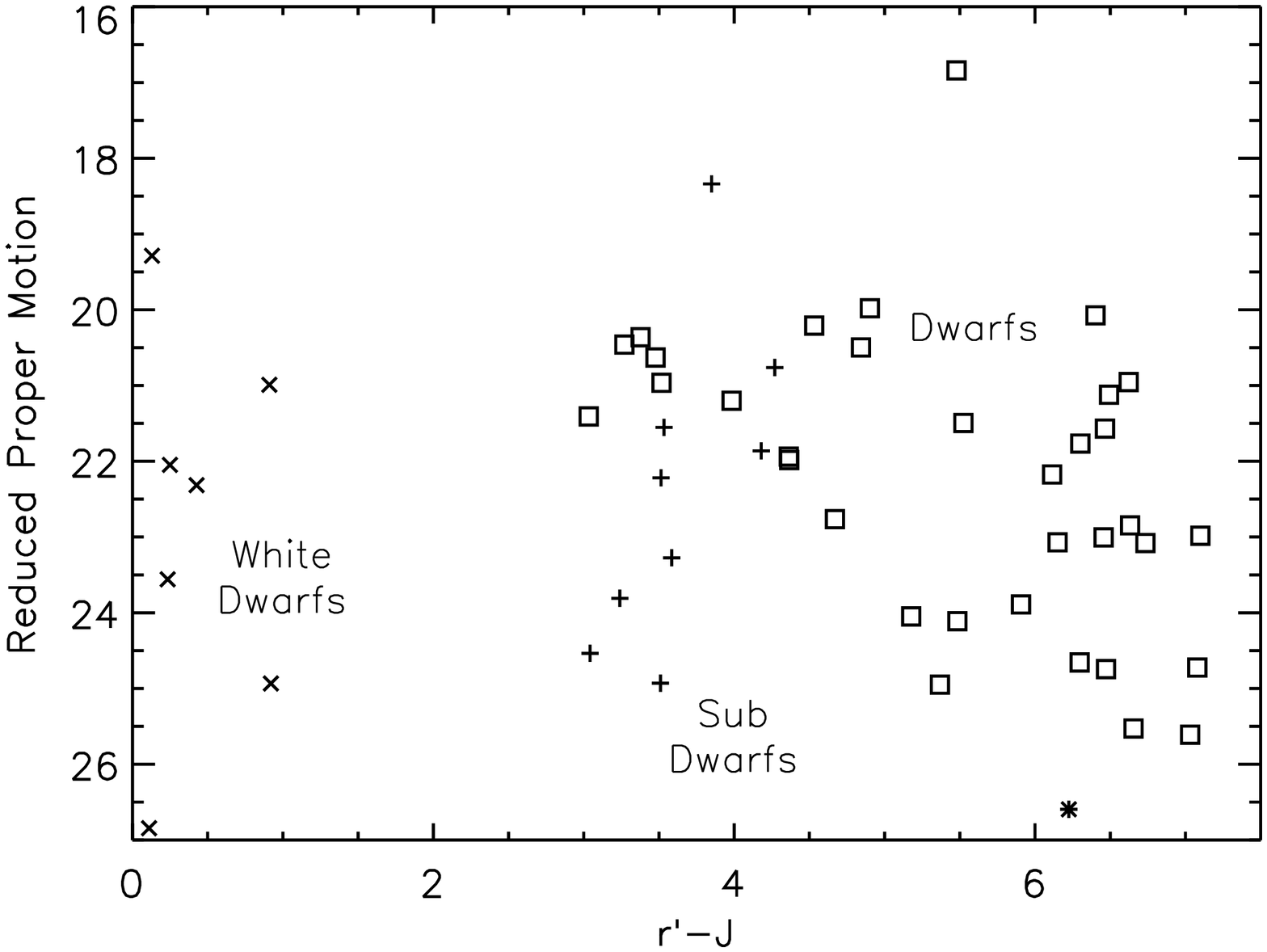}}
\caption{The $r'-J$ colors versus the reduced proper motion (RPM) for
the 36 newly indentified high proper motion objects (squares), white
dwarfs discovered in this survey (x's), known M subdwarfs (plus signs)
as found in the SDSS and confirmed by Marshall (2008) and Lepine and
Scholz (2008) and the known L4 subdwarf 2MASS J1626$+$3925 (asterisk)
that was detected in this survey and identified as a subdwarf by
Burgasser et al. (2007).  The white dwarfs occupy the left portion of
the figure and none of the newly identified HPM objects are near this
region.  The subdwarfs occupy the center and lower portions of the
figure.  Several of the newly identified HPM objects are near the
subdwarf region and thus are good subdwarf candidates (see Table 4).
2MASS J1434+2202 was identified as a possible L subdwarf from followup
spectroscopy and is the bottom most point at the far right of this
figure near the known L subdwarf 2MASS J1626$+$3925.  Most of the
newly identified HPM objects are in the far right of this figure which
indicates that most are dwarf type objects.  The one T dwarf
discovered in this survey is not shown since its $r'$ magnitude is
unknown.}
\label{fig:reducedPM} 
\end{figure}

In Figure \ref{fig:reducedPM} we plot the $r'-J$ colors versus the RPM
for the newly discovered high proper motion objects discovered in this
survey along with some white dwarfs found in this survey, the L4
subdwarf 2MASS J1626$+$3925 also found in this survey and several
confirmed M subdwarfs seen in both the SDSS and 2MASS from Marshall
(2008) and Lepine and Scholz (2008).  It can clearly be seen that the
white dwarfs occupy the far left in the figure while the subdwarfs
occupy the center and lower portions of Figure \ref{fig:reducedPM}.
Several of our newly identified HPM objects are near the center and
lower portions of Figure \ref{fig:reducedPM} and are thus good
subdwarf candidates (see Table 4).  In fact, 2MASS J1434$+$2202, which
is near the known subdwarf 2MASS J1626$+$3925 in Figure
\ref{fig:reducedPM}, has been identified as a likely subdwarf from our
near-infrared spectroscopy (see Table 7).  Most of our newly found HPM
objects are on the far right of Figure \ref{fig:reducedPM} and thus
not likely subdwarfs and are definitely not white dwarfs.  Because the
scatter is rather large a more detailed spectroscopic observational
campaign will be required to determine which of these objects are
subdwarfs.

\section{Summary}

About 8000 square degrees from the 2MASS and SDSS catalogs were
compared to find low mass (red), high proper motion objects.
Thirty-six new HPM objects were discovered in the analysis.  All
objects were confirmed as having high proper motions with deep imaging
at the du Pont 2.5 meter and Magellan 6.5 meter telescopes.  Their red
optical and near infrared colors indicate that 27 are M dwarfs, 8 are
early-type L dwarfs, and 1 is a late-type T dwarf.  Through comparing
the colors and reduced proper motions of our newly identified HPM
objects we find several are good subdwarf candidates.

The newly identified L dwarfs appear to have $J-K_s$ colors at the
extreme of known L dwarf colors.  This may hint at a slight color bias
in brown dwarf detections through color surveys.  Followup
near-infrared spectroscopy of six of the late-type objects show that
2MASS J14313097$+$1436539 and 2MASS J14343616$+$2202463 exhibit
peculiar spectral features similar to other peculiar ``blue'' L
dwarfs.  Spectroscopy also shows that 2MASS 14343616$+$2202463 could
be an L subdwarf and 2MASS 10595185$+$3042059 is a T dwarf.  From
their estimated absolute magnitudes it appears all of the brown dwarfs
found in this survey are within 100 parsecs of the Earth.

In addition, several hundred previously known high proper motion
objects were also detected as well as ten known L- and five known
T-dwarfs.  For the first time we determined the proper motions and
position angles for known T-dwarfs SDSS J085834.42$+$325627.6 and SDSS
J125011.65$+$392553.9 as well as the late L-dwarf 2MASS
J15261405$+$2043414.

An additional 13 objects that appeared real in the 2MASS catalog but are
not identified or correlated with any objects in the SDSS were found.
Deep du Pont 2.5 meter $I$-band imaging of these locations found no
interesting counterparts.  These objects are likely artifacts in the
2MASS catalog but could be extremely red objects that could not be
detected in the shorter wavelength SDSS images or very fast moving
objects of well over 10$''$ per year.

\section*{Acknowledgments}

We thank Alan Tokunaga for granting us director's discretionary time on
the IRTF.  This research has benefitted from the M, L, and T dwarf
compendium housed at DwarfArchives.org and maintained by Chris Gelino,
Davy Kirkpatrick, and Adam Burgasser and the SpeX Prism Spectral
Libraries, maintained by Adam Burgasser at
http://www.browndwarfs.org/spexprism.  This paper includes data gathered
with the 6.5 meter Magellan Telescopes located at Las Campanas
Observatory, Chile.  This publication makes use of data products from
the Two Micron All Sky Survey, which is a joint project of the
University of Massachusetts and the Infrared Processing and Analysis
Center/California Institute of Technology, funded by the National
Aeronautics and Space Administration and the National Science
Foundation.  The SDSS is managed by the Astrophysical Research
Consortium for the Participating Institutions. The Participating
Institutions are the American Museum of Natural History, Astrophysical
Institute Potsdam, University of Basel, Cambridge University, Case
Western Reserve University, University of Chicago, Drexel University,
Fermilab, the Institute for Advanced Study, the Japan Participation
Group, Johns Hopkins University, the Joint Institute for Nuclear
Astrophysics, the Kavli Institute for Particle Astrophysics and
Cosmology, the Korean Scientist Group, the Chinese Academy of Sciences
(LAMOST), Los Alamos National Laboratory, the Max-Planck-Institute for
Astronomy (MPIA), the Max-Planck-Institute for Astrophysics (MPA), New
Mexico State University, Ohio State University, University of
Pittsburgh, University of Portsmouth, Princeton University, the United
States Naval Observatory, and the University of Washington.  Funding for
the SDSS and SDSS-II has been provided by the Alfred P. Sloan
Foundation, the Participating Institutions, the National Science
Foundation, the U.S. Department of Energy, the National Aeronautics and
Space Administration, the Japanese Monbukagakusho, the Max Planck
Society, and the Higher Education Funding Council for England.  The SDSS
web site is http://www.sdss.org.

%{\it Facilities:} 
%\facility{Magellan:Clay (LDSS3), IRTF (SPEX)}

\newpage

% This is a template LaTeX input file.  (Version of 17 August 1999)
% 

%\documentstyle [aj_pt4]{article}    % Specifies the document style.

%\begin{document}

%\begin{center}
\begin{deluxetable}{lccccccc}
\tabletypesize{\scriptsize}
\tablenum{1}
\tablewidth{0pc}
\tablecaption{\label{tab:tab0} Properties of Known L and T Dwarfs Detected in this Survey}
\tablecolumns{8}
\tablehead{ \colhead{Name} & \colhead{PM\tablenotemark{a}} & \colhead{PA\tablenotemark{b}} & \colhead{$J$\tablenotemark{c}} & \colhead{$i'-J$} & \colhead{Spec.\tablenotemark{d}} & \colhead{Ref.\tablenotemark{e}} \\ \colhead{} & \colhead{($\arcsec /$yr)} & \colhead{(deg)}  & \colhead{(mag)} & \colhead{(mag)} & \colhead{Type} & \colhead{}}
\startdata
2MASS J14213145+1827407   & 0.79                   & 257                   & 13.23  & 4.2      & L0   & 9,10,17 \nl
2MASS J14392836+1929149   & 1.34                   & 288                   & 12.76  & 4.4      & L1   & 2,12 \nl
2MASS J13004255+1912354   & 1.48                   & 213                   & 12.72  & 4.5      & L1   & 9,10,11 \nl 
2MASS J15065441+1321060   & 1.14                   & 269                   & 13.37  & 4.6      & L3   & 9,13 \nl
2MASS J16154416+3559005   & 0.53                   & 184                   & 14.54  & 4.6      & L3   & 1,13 \nl
2MASS J16262034+3925190   & 1.79                   & 279                   & 14.44  & 3.5      & sdL4 & 8,16 \nl
2MASS J13285503+2114486   & 0.50                   & 150                   & 16.20  & 5.2      & L5   & 2,12 \nl
2MASS J15150083+4847416   & 2.04                   & 328                   & 14.11  & 5.2      & L6   & 13,14,15 \nl
2MASS J15261405+2043414   & 0.43\tablenotemark{*}  & 216\tablenotemark{*}  & 15.59  & 5.0      & L7   & 1 \nl
2MASS J08251968+2115521   & 0.62                   & 241                   & 15.10  & 5.5      & L7.5 & 1,2,3 \nl
SDSS J085834.42+325627.6  & 0.76\tablenotemark{*}  & 272\tablenotemark{*}  & 16.45  & 5.6      & T1   & 4 \nl
SDSS J125011.65+392553.9  & 1.03\tablenotemark{*}  & 182\tablenotemark{*}  & 16.54  & $>5.5$   & T4   & 4 \nl
2MASS J12314753+0847331   & 1.51                   & 228                   & 15.57  & $>6.4$   & T5.5 & 6,8 \nl
2MASS J09373487+2931409   & 1.68                   & 143                   & 14.65  & $>7.3$   & T6p  & 5,6,7 \nl
2MASS J15530228+1532369   & 0.44                   & 294                   & 15.83  & $>6.2$   & T7   & 5,6,13 \nl
\enddata
\tablenotetext{*}{Indicates this is the first time that this type of measurement for this object has been published.}
\tablenotetext{a}{The Proper Motion (PM) of the object as observed between the 2MASS and SDSS data.  Uncertainties are at about the 10 percent level.}
\tablenotetext{b}{The Position Angle (PA) of the motion.  Uncertainties are generally about 1 degree.}
\tablenotetext{c}{The $J$-band photometry from 2MASS. Uncertainties are less than 0.1 magnitudes.}
\tablenotetext{d}{The spectral types of the L dwarfs are based on red optical spectroscopy and are on the Kirkpatrick et al. (1999) system.  The spectral types of the T dwarfs are based on near-infrared spectoscopy and are on the Burgasser et al. (2006) system.  The spectral type for 2MASS J16262034$+$3925190 is tentative (Burgasser et al.  2007a).}
\tablenotetext{e}{References for discovery, spectral type identification, proper motion and position angle measurements if applicable:}
\tablerefs{1) Kirkpatrick et al. (2000), 2) Dahn et al. (2002), 3) Knapp et al. (2004), 4) Chiu et al. (2006), 5) Burgasser et al. (2002), 6) Burgasser et al. (2006), 7) Vrba et al. (2004), 8) Burgasser et al. (2004a), 9) Gizis et al. (2000), 10) Schmidt et al. (2007), 11) Burgasser et al. (2008), 12) Kirkpatrick et al. (1999), 13) Jameson et al. (2008), 14) Wilson et al. (2003), 15) Cruz et al. (2007), 16) Burgasser, Cruz, and Kirkpatrick (2007), 17) Reid et al. in preparation.}
\end{deluxetable}
%\end{center}

%\end{document}             % End of document.

\newpage

% This is a template LaTeX input file.  (Version of 17 August 1999)
% 

%\documentstyle [aj_pt4]{article}    % Specifies the document style.

%\begin{document}

%\begin{center}
\begin{deluxetable}{lccccc}
\tabletypesize{\scriptsize}
\tablenum{2}
\tablewidth{0pc}
\tablecaption{\label{tab:tab2} Objects in the 2MASS PSC but not in the SDSS PSC\tablenotemark{a}}
\tablecolumns{6}
\tablehead{
\colhead{2MASS Name} & \colhead{$J$\tablenotemark{b}} & \colhead{$H$\tablenotemark{b}} & \colhead{$K_s$\tablenotemark{b}} & \colhead{JD\tablenotemark{d}} & \colhead{Comments\tablenotemark{c}} \\ \colhead{} & \colhead{(mag)} & \colhead{(mag)} & \colhead{(mag)} & \colhead{} & \colhead{} }
\startdata
J00461651+1419298 & 16.9  & 16.3 & 15.9 & 2451437.8263 & Diff \nl
J08204794+2440364 & 16.9  & 16.2 & 15.8 & 2451506.9024 & Diff \nl
J09581895+0143068 & 16.8  & 16.8 & 16.1 & 2451577.7483 & Noise \nl
J10385877+2718553 & 16.9  & 16.3 & 15.7 & 2450873.8665 & Diff \nl
J11254075+5029579 & 16.5  & 16.2 & 15.8 & 2451601.7506 & Diff \nl
J11280132+0715387 & 16.9  & 17.2 & 15.9 & 2451605.6786 & Noise \nl
J13165176+1338169 & 16.8  & 16.4 & 16.3 & 2451526.0489 & Noise \nl
J13300094+4912080 & 16.9  & 16.8 & 16.4 & 2451621.8835 & Noise \nl
J14510138+2847574 & 16.5  & 16.1 & 15.3 & 2451335.6506 & Diff \nl
J15490517+1542418 & 16.9  & 16.6 & 16.0 & 2451557.0605 & Noise \nl
J16382286+2501307 & 16.9  & 16.5 & 16.2 & 2451621.9096 & Noise \nl
J17262351+2708338 & 16.6  & 16.3 & 15.5 & 2451628.9394 & Diff \nl
J23390347+1415329 & 16.9  & 16.4 & 16.1 & 2451877.6576 & Noise \nl
\enddata
\tablenotetext{a}{These are objects that were not linked to known asteroids, not seen in the SDSS and were not obvious artifacts or noise.  These objects were cataloged by 2MASS in $J$, $H$ and $K_s$.  The above objects are all near the detection limit of 2MASS.  Most of the objects are likely artifacts from scattered light of nearby bright stars but some could be unknown asteroids, extremely red objects ($i'-J>7$ mags) or extremely high proper motion stars or brown dwarfs ($>> 10\arcsec$ yr$^{-1}$).  Deep $I$-band imaging of these areas with the Dupont 2.5 meter telescope found no objects near any of the coordinates listed above.}
\tablenotetext{b}{The $J$, $H$ and $K_s$ photometry from 2MASS.  Most of these objects are near the limit of 2MASS detection and thus their uncertainties are higher than most 2MASS photometry and are generally between 0.2 and 0.4 magnitudes.}
\tablenotetext{c}{Most of these objects are likely artifacts in the 2MASS data and here we comment on what they likely are: Diff $=$ Differential diffration spike from nearby bright star or Noise = Background Noise.}
\tablenotetext{d}{The Julian Date of the 2MASS image and coordinates.}
\end{deluxetable}
%\end{center}

%\end{document}             % End of document.

\newpage

% This is a template LaTeX input file.  (Version of 17 August 1999)
% 

%\documentstyle [aj_pt4]{article}    % Specifies the document style.

%\begin{document}

\begin{center}
\begin{deluxetable}{lccccccccc}
\tabletypesize{\scriptsize}
\tablenum{3}
\tablewidth{0pc}
\tablecaption{\label{tab:tab1} New High Proper Motion Objects\tablenotemark{a}}
\tablecolumns{10}
\tablehead{
\colhead{2MASS Name} & \colhead{PM\tablenotemark{b}} & \colhead{PA\tablenotemark{c}} & \colhead{$r'$\tablenotemark{d}} & \colhead{$i'$\tablenotemark{d}} & \colhead{$z'$\tablenotemark{d}} & \colhead{$J$\tablenotemark{e}} & \colhead{$H$\tablenotemark{e}} & \colhead{$K_s$\tablenotemark{e}} & \colhead{JD\tablenotemark{f}} \\ \colhead{} & \colhead{($\arcsec /$yr)} & \colhead{(deg)}  & \colhead{(mag)} & \colhead{(mag)} & \colhead{(mag)} & \colhead{(mag)} & \colhead{(mag)} & \colhead{(mag)} & \colhead{} }
\startdata
2MASS J08404809+2024121 & 0.43  &   121 & 22.2 & 19.2 & 17.6 & $15.78\pm0.06 $ & $15.16\pm0.08 $ & $14.7\pm0.1 $ & 2451106.0059 \nl
2MASS J09122241+2258119 & 0.36  &   221 & 20.1 & 17.9 & 16.7 & $15.22\pm0.04 $ & $14.78\pm0.05 $ & $14.4\pm0.1 $ & 2450825.8190 \nl
2MASS J09585080+2001525 & 0.38  &   223 & 22.7 & 20.5 & 18.6 & $16.08\pm0.08 $ & $15.25\pm0.08 $ & $14.6\pm0.1 $ & 2450838.7923 \nl
2MASS J10001708+3218306 & 0.45  &   240 & 23.7 & 21.0 & 18.7 & $16.62\pm0.12 $ & $15.70\pm0.11 $ & $15.4\pm0.2 $ & 2450893.7960 \nl
2MASS J10120045+2046128 & 0.44  &   242 & 19.7 & 18.3 & 17.5 & $16.21\pm0.07 $ & $15.69\pm0.09 $ & $15.7\pm0.2 $ & 2450839.8136 \nl
2MASS J10182596+2710554 & 0.35  &   177 & 21.3 & 18.2 & 16.6 & $14.81\pm0.03 $ & $14.30\pm0.05 $ & $14.0\pm0.1 $ & 2450868.8630 \nl
2MASS J10340305+2008184 & 0.34  &   189 & 20.4 & 17.5 & 15.9 & $14.00\pm0.02 $ & $13.40\pm0.03 $ & $13.0\pm0.0 $ & 2450841.8812 \nl
2MASS J10351535+2736016 & 0.41  &   153 & 19.8 & 18.5 & 17.8 & $16.45\pm0.10 $ & $15.92\pm0.13 $ & $15.8\pm0.2 $ & 2450873.8263 \nl
2MASS J10512777+1816421 & 0.40  &   144 & 20.0 & 18.7 & 17.9 & $16.72\pm0.14 $ & $15.71\pm0.14 $ & $15.8\pm0.2 $ & 2450842.9695 \nl
2MASS J10543739+1426556 & 0.68  &   252 & 21.9 & 19.3 & 18.0 & $16.48\pm0.12 $ & $15.85\pm0.14 $ & $15.9\pm0.2 $ & 2451261.7513 \nl
2MASS J10595185+3042059 & 0.53  &   184 & $\cdots$ & 24.0 & 19.8 & $16.21\pm0.09 $ & $15.80\pm0.12 $ & $15.6\pm0.2 $ & 2450893.8534 \nl
2MASS J11092121+4425488 & 0.73  &   234 & 18.0 & 16.8 & 16.3 & $14.97\pm0.05 $ & $14.50\pm0.04 $ & $14.5\pm0.1 $ & 2451266.7371 \nl
2MASS J11101471+1731121 & 0.37  &   269 & 21.5 & 18.6 & 17.1 & $15.02\pm0.03 $ & $14.32\pm0.03 $ & $14.0\pm0.0 $ & 2450800.0062 \nl
2MASS J11212384+2018192 & 0.68  &   246 & 21.7 & 18.7 & 17.1 & $15.20\pm0.04 $ & $14.64\pm0.05 $ & $14.2\pm0.1 $ & 2450846.8717 \nl
2MASS J11261886+4429311 & 0.57  &   167 & 20.6 & 18.5 & 17.4 & $15.93\pm0.06 $ & $15.40\pm0.08 $ & $15.6\pm0.1 $ & 2451247.7611 \nl
2MASS J11303803+2341480 & 0.57  &   256 & 23.3 & 21.3 & 19.1 & $16.65\pm0.13 $ & $15.87\pm0.12 $ & $15.7\pm0.2 $ & 2451525.9601 \nl
2MASS J11355548+2639401 & 0.64  &   229 & 19.2 & 17.3 & 16.3 & $14.84\pm0.04 $ & $14.31\pm0.04 $ & $14.0\pm0.1 $ & 2451579.8211 \nl
2MASS J11470670+1729029 & 0.38  &   252 & 20.8 & 18.1 & 16.4 & $14.22\pm0.03 $ & $13.57\pm0.03 $ & $13.1\pm0.0 $ & 2450822.9444 \nl
2MASS J11541839+2248569 & 0.44  &   166 & 20.3 & 18.6 & 17.7 & $16.30\pm0.09 $ & $15.60\pm0.11 $ & $15.1\pm0.1 $ & 2451260.8033 \nl
2MASS J12205687+2132371 & 0.51  &   170 & 20.6 & 17.8 & 16.3 & $14.47\pm0.03 $ & $13.97\pm0.03 $ & $13.6\pm0.0 $ & 2450948.6895 \nl
2MASS J12304562+2827583 & 0.58  &   263 & 22.4 & 20.1 & 18.5 & $16.07\pm0.09 $ & $15.00\pm0.08 $ & $14.4\pm0.1 $ & 2451653.7353 \nl
2MASS J12353360+1805024 & 0.47  &   281 & 19.3 & 17.1 & 15.9 & $14.41\pm0.03 $ & $13.91\pm0.04 $ & $13.7\pm0.0 $ & 2451293.6837 \nl
2MASS J12370037+2107445 & 0.54  &   273 & 19.9 & 17.1 & 15.6 & $13.58\pm0.03 $ & $12.96\pm0.02 $ & $12.6\pm0.0 $ & 2451293.6962 \nl
2MASS J12373441+3028596 & 0.74  &   313 & 23.8 & 20.5 & 18.6 & $16.37\pm0.14 $ & $16.27\pm0.31 $ & $15.5\pm0.2 $ & 2450876.9792 \nl
2MASS J12531161+2728145 & 0.53  &   151 & 21.2 & 18.4 & 16.6 & $14.49\pm0.04 $ & $13.84\pm0.04 $ & $13.3\pm0.0 $ & 2451555.9499 \nl
2MASS J12572875+1555356 & 0.58  &   160 & 19.7 & 17.8 & 16.8 & $15.29\pm0.04 $ & $14.87\pm0.06 $ & $14.5\pm0.1 $ & 2450836.9610 \nl
2MASS J13001921+2036407 & 0.50  &   289 & 20.0 & 17.5 & 16.3 & $14.47\pm0.03 $ & $13.89\pm0.03 $ & $13.5\pm0.0 $ & 2451261.8390 \nl
2MASS J13034203+2158519 & 0.87  &   179 & 19.8 & 17.3 & 16.0 & $14.33\pm0.03 $ & $13.79\pm0.03 $ & $13.6\pm0.0 $ & 2451261.8670 \nl 
2MASS J13512249+1419168 & 0.60  &   246 & 21.6 & 19.1 & 18.0 & $16.43\pm0.10 $ & $16.07\pm0.19 $ & $15.7\pm0.2 $ & 2450935.7796 \nl
2MASS J13580384+1458204 & 0.51  &   298 & 22.3 & 20.5 & 18.9 & $16.37\pm0.11 $ & $15.25\pm0.11 $ & $14.7\pm0.1 $ & 2451531.0524 \nl
2MASS J14005107+2142014 & 0.41  &   193 & 19.7 & 17.8 & 16.8 & $15.18\pm0.05 $ & $14.83\pm0.08 $ & $14.5\pm0.1 $ & 2450953.8086 \nl
2MASS J14222455+2834542 & 0.56  &   261 & 21.0 & 18.1 & 16.7 & $14.83\pm0.04 $ & $14.36\pm0.04 $ & $14.0\pm0.1 $ & 2451264.9276 \nl
2MASS J14313097+1436539 & 0.43  &   259 & 22.3 & 19.7 & 17.6 & $15.15\pm0.04 $ & $14.50\pm0.05 $ & $14.1\pm0.1 $ & 2450935.8401 \nl
2MASS J14343616+2202463 & 0.83  &   286 & 21.6 & 18.8 & 16.8 & $14.52\pm0.04 $ & $13.83\pm0.04 $ & $13.6\pm0.0 $ & 2451557.0310 \nl
2MASS J15052821+3115037 & 0.53  &   184 & 19.1 & 17.7 & 17.0 & $15.58\pm0.15 $ & $14.99\pm0.07 $ & $14.8\pm0.1 $ & 2450897.9287 \nl
2MASS J17143190+3544160 & 0.13  &   345 & 21.9 & 19.6 & 18.2 & $16.42\pm0.11 $ & $16.15\pm0.21 $ & $15.7\pm0.2 $ & 2450913.9061 \nl
\enddata
\tablenotetext{a}{Objects that were found in both 2MASS and SDSS catalogs, moved $>2.35$ arcseconds and were not in the SIMBAD database or known brown dwarfs on the dwarf archives page (dwarfarchives.org) as of May 1, 2008. 2MASS astrometry is good to about 80 mas (Skrutskie et al. 2006).}
\tablenotetext{b}{The Proper Motion (PM) of the object as observed between the 2MASS and SDSS data.  Uncertainties are at about the 10 percent level.}
\tablenotetext{c}{The Position Angle (PA) of the objects motion.  Uncertainties are generally about 1 degree.}
\tablenotetext{d}{The r', i' and z' photometry from the SDSS.  Uncertainties are generally well below 0.1 magnitudes. The only exception is the i' photometry for 2MASS J10595185+3042059.  This object was not seen in the i' SDSS data and was only marginally detected in one 200 second image using LDSS3 on the 6.5 meter Clay telescope.  Thus the uncertainty in the i'-band photometry for 2MASS J10595185+3042059 is around 0.3 magnitudes.}
\tablenotetext{e}{The J, H and K photometry from 2MASS.}
\tablenotetext{f}{The Julian Date (JD) of the 2MASS image and coordinates.}
\end{deluxetable}
\end{center}

%\end{document}             % End of document.

\newpage

% This is a template LaTeX input file.  (Version of 17 August 1999)
% 

%\documentstyle [aj_pt4]{article}    % Specifies the document style.

%\begin{document}

%\begin{center}
\begin{deluxetable}{lccccccc}
\tabletypesize{\scriptsize}
\tablenum{4}
\tablewidth{0pc}
\tablecaption{\label{tab:tab3} Colors and Distances of Newly Identified High Proper Motion Objects}
\tablecolumns{8}
\tablehead{
\colhead{2MASS Name} & \colhead{$r'-i'$} & \colhead{$i'-J$} & \colhead{$J-K_s$} & \colhead{RPM\tablenotemark{a}} & \colhead{M$_{J}$\tablenotemark{b}} & \colhead{D\tablenotemark{c}} & \colhead{Spec\tablenotemark{d}} \\ \colhead{} & \colhead{(mag)} & \colhead{(mag)} & \colhead{(mag)} & \colhead{} & \colhead{(mag)} & \colhead{(pc)} & \colhead{Type}}
\startdata
2MASS J11092121+4425488  &    $1.2\pm0.1$  &    $1.83\pm0.05$  &   $0.50\pm0.08$ &  21.4*   & 6.5*  & 500* &   M0 \nl
2MASS J10512777+1816421  &    $1.3\pm0.1$  &    $1.93\pm0.15$  &   $0.92\pm0.27$ &  20.5*   & 7.0* &  860* &   M1 \nl
2MASS J10351535+2736016  &    $1.4\pm0.1$  &    $2.03\pm0.11$  &   $0.65\pm0.21$ &  20.4*   & 7.3* &  90*  &   M1 \nl
2MASS J10120045+2046128  &    $1.4\pm0.1$  &    $2.06\pm0.09$  &   $0.52\pm0.19$ &  20.6*   & 7.5* &  550* &   M1 \nl
2MASS J15052821+3115037  &    $1.4\pm0.1$  &    $2.10\pm0.05$  &   $0.80\pm0.11$ &  21.0*   & 7.7*  & 400* &   M1 \nl
2MASS J11541839+2248569  &    $1.7\pm0.1$  &    $2.28\pm0.10$  &   $1.16\pm0.15$ &  21.2*   & 8.2* &  60*  &   M2 \nl
2MASS J11355548+2639401  &    $1.9\pm0.1$  &    $2.45\pm0.05$  &   $0.83\pm0.06$ &  22.0   & 8.7 &  170 &   M3 \nl
2MASS J12572875+1555356  &    $1.9\pm0.1$  &    $2.46\pm0.05$  &   $0.78\pm0.08$ &  21.9   & 8.7 &  205 &   M3 \nl
2MASS J14005107+2142014  &    $2.0\pm0.1$  &    $2.57\pm0.07$  &   $0.68\pm0.17$ &  20.2   & 9.0 &  170 &   M3 \nl
2MASS J11261886+4429311  &    $2.1\pm0.1$  &    $2.58\pm0.05$  &   $0.30\pm0.11$ &  22.8   & 9.1  & 230 &   M3 \nl
2MASS J12353360+1805024  &    $2.2\pm0.1$  &    $2.64\pm0.04$  &   $0.72\pm0.05$ &  20.5   & 9.2 &  110 &   M3 \nl
2MASS J09122241+2258119  &    $2.2\pm0.1$  &    $2.68\pm0.06$  &   $0.83\pm0.06$ &  20.0   & 9.3 &  155 &   M4 \nl
2MASS J13512249+1419168  &    $2.5\pm0.1$  &    $2.68\pm0.10$  &   $0.75\pm0.26$ &  24.0   & 9.2 &  275 &   M4 \nl
2MASS J10543739+1426556  &    $2.6\pm0.1$  &    $2.80\pm0.14$  &   $0.62\pm0.28$ &  25.0*   & 9.5* &  250* &   M5 \nl
2MASS J13034203+2158519  &    $2.6\pm0.1$  &    $2.92\pm0.04$  &   $0.76\pm0.04$ &  24.1   & 9.8 &  80  &   M5 \nl
2MASS J13001921+2036407  &    $2.5\pm0.1$  &    $3.03\pm0.04$  &   $0.95\pm0.05$ &  21.5   & 10.0 & 80  &   M5 \nl
2MASS J17143190+3544160  &    $2.3\pm0.1$  &    $3.20\pm0.14$  &   $0.70\pm0.25$ &  16.8   & 10.2 & 170 &   M6 \nl
2MASS J12205687+2132371  &    $2.8\pm0.1$  &    $3.28\pm0.04$  &   $0.88\pm0.05$ &  22.2   & 10.4 & 65  &   M6 \nl
2MASS J14222455+2834542  &    $2.9\pm0.1$  &    $3.30\pm0.04$  &   $0.82\pm0.06$ &  23.1   & 10.4 & 75  &   M7 \nl
2MASS J10182596+2710554  &    $3.1\pm0.1$  &    $3.37\pm0.06$  &   $0.84\pm0.06$ &  21.1   & 10.5 & 70  &   M7 \nl
2MASS J08404809+2024121  &    $3.1\pm0.2$  &    $3.39\pm0.09$  &   $1.05\pm0.13$ &  23.0   & 10.6 & 110 &   M8 \nl
2MASS J12370037+2107445  &    $2.8\pm0.1$  &    $3.49\pm0.04$  &   $1.02\pm0.04$ &  21.8   & 10.7 & 40  &   M8 \nl
2MASS J11212384+2018192  &    $3.0\pm0.1$  &    $3.50\pm0.07$  &   $1.00\pm0.07$ &  24.7   & 10.7 & 80  &   M8 \nl
2MASS J10340305+2008184  &    $2.9\pm0.1$  &    $3.52\pm0.05$  &   $1.00\pm0.04$ &  20.1   & 10.8 & 45  &   M8 \nl
2MASS J11101471+1731121  &    $2.9\pm0.1$  &    $3.56\pm0.06$  &   $1.07\pm0.05$ &  21.6   & 10.8 & 70  &   M8 \nl
2MASS J11470670+1729029  &    $2.7\pm0.1$  &    $3.89\pm0.05$  &   $1.12\pm0.04$ &  21.0   & 11.3 & 40  &   M9 \nl
2MASS J12531161+2728145  &    $2.8\pm0.1$  &    $3.91\pm0.06$  &   $1.14\pm0.05$ &  23.1   & 11.4 & 40  &   M9 \nl
2MASS J12304562+2827583  &    $2.3\pm0.2$  &    $4.02\pm0.10$  &   $1.63\pm0.11$ &  24.7   & 11.5 & 80  &   L0 \nl
2MASS J12373441+3028596  &    $3.3\pm0.3$  &    $4.10\pm0.11$  &   $0.90\pm0.25$ &  27.3   & 11.6 & 90  &   L0 \nl
2MASS J13580384+1458204  &    $1.8\pm0.2$  &    $4.10\pm0.11$  &   $1.71\pm0.14$ &  23.9   & 11.6 & 90  &   L0 \nl
2MASS J14343616+2202463  &    $2.8\pm0.1$  &    $4.28\pm0.04$  &   $0.97\pm0.06$ &  25.6*   & 11.9* & 35*  &   L1 \nl
2MASS J09585080+2001525  &    $2.3\pm0.2$  &    $4.38\pm0.11$  &   $1.46\pm0.11$ &  22.9   & 12.1 & 65  &   L2 \nl
2MASS J10001708+3218306  &    $2.7\pm0.6$  &    $4.40\pm0.15$  &   $1.20\pm0.14$ &  24.7   & 12.1 & 80  &   L2 \nl
2MASS J14313097+1436539  &    $2.6\pm0.1$  &    $4.54\pm0.04$  &   $1.02\pm0.07$ &  23.0   & 12.3 & 40  &   L2 \nl
2MASS J11303803+2341480  &    $2.0\pm0.2$  &    $4.61\pm0.15$  &   $0.90\pm0.22$ &  25.5*   & 12.4* & 70*  &   L3 \nl
2MASS J10595185+3042059  &    $\cdots$     &    $7.80\pm0.31$  &   $0.67\pm0.21$ &  31.8   & 14.1 & 25  &   T8 \nl  
\enddata
\tablenotetext{a}{The Reduced Proper Motion (RPM) of the objects.  An object with an * indicates it is a good subdwarf candidate (see Figure 9).  If it is a subdwarf than the absolute magnitude and distance estimates are unreliable.}
\tablenotetext{b}{Approximate absolute magnitude based on the $i'-J$
  color information as described in Hawley et al. (2002).  Uncertainties
  are a few tenths of magnitudes based on comparing the results for known objects from Table 1.  An object with an * indicates it is a good subdwarf candidate (see Figure 9).  If it is a subdwarf than the absolute magnitude and distance estimates are unreliable.}  
\tablenotetext{c}{Estimated distances for the objects are good to about the 25 percent level based on comparing the results to known objects from Table 1.  An object with an * indicates it is a good subdwarf candidate (see Figure 9).  If it is a subdwarf than the absolute magnitude and distance estimates are unreliable.}
\tablenotetext{d}{The spectral type is estimated based on the $i'-J$
  colors as shown in Figure 5.  The estimated spectral type based on the
  $i'-J$ colors is likely accurate to within a subclass except for 2MASS 1059+3042 which is good to within about 3 subclasses since the $z'-J$ colors were used to find the spectral type for this extremely red object.}
\end{deluxetable}
%\end{center}

%\end{document}             % End of document.

\newpage

\begin{deluxetable}{llcl}
\tablenum{5}
\tablecolumns{4}
\tabletypesize{\scriptsize}
\tablewidth{0pc}
\tablecaption{\label{tab:SpeXObsLog}Log of SpeX Observations}
\tablehead{
\colhead{Object} & 
\colhead{UT Date} & 
\colhead{Exp. Time (sec)} & 
\colhead{A0 V}}

\startdata
2MASS J1059+3042 & 2008 Jun 13 & 1200    & HD 89239 \\
2MASS J1230+2827 & 2008 Jun 18 & 1440    & HD 107655 \\
2MASS J1237+3028 & 2008 Jun 18 & 1200    & HD 107655 \\
2MASS J1253+2728 & 2008 Jun 18 & 1200    & HD 107655 \\
2MASS J1431+1436 & 2005 Jun 18 & 1200    & HD 121880 \\ 
2MASS J1434+2202 & 2008 Jun 18 & \phn960 & HD 121880 \\

\enddata
\end{deluxetable}

\newpage

\begin{deluxetable}{llllll}
  \tablecolumns{6}
\tablenum{6}
\tabletypesize{\scriptsize} 
\tablewidth{0pc}
\tablecaption{\label{tab:Indices}Spectral Indices of Late M and Early L Dwarfs}
\tablehead{
\colhead{} & 
\multicolumn{5}{c}{Spectral Type} \\
\cline{2-6}

\colhead{Index} &
\colhead{2MASS 1230} &
\colhead{2MASS 1237} &
\colhead{2MASS 1253} &
\colhead{2MASS 1431} &
\colhead{2MASS 1434}}

\startdata
 
\multicolumn{6}{c}{Reid et al. (2001) }\\
\hline
H$_2$O$^A$         & M9 V   & M7.5 V & L0.5    & L5.5  & L3.5 \\
H$_2$O$^B$         & L1.5   & M8 V   & L0      & L4.5  & L3.5 \\

\hline
\multicolumn{6}{c}{Testi et al. (2001) }\\
\hline

%sHJ                & L3     & $<$L0  & $<$L0   & $<$L0 & $<$L0 \\
%sKJ                & L3.5   & $<$L0  & $<$L0   & $<$L0 & $<$L0 \\
sH$_2$O$^J$        & L3     & $<$L0  & L0      & L1.5  & L0.5 \\
sH$_2$O$^{H1}$     & L1.5   & $<$L0  & L0      & L4    & L2.5 \\
sH$_2$O$^{H2}$     & $<$L0  & L0     & L2      & L5.5  & L5 \\
sH$_2$O$^K$        & L1     & $<$L0  & $<$L0   & L2.5  & L1.5 \\

\hline
\multicolumn{6}{c}{Burgasser et al. (2002) }\\
\hline

H$_2$O-A           & L1.5   & M6.5 V & M8 V    & L2    & L0.5 \\
H$_2$O-B           & M8.5 V & M6.5 V & M8.5 V  & L6    & L4 \\
H$_2$O-C           & L2     & M5  V  & M8 V    & L5.5  & L2 \\
%H/J                & L5.5   & M6.5 V & M6.5 V  & L4    & L2.5 \\
%K/J                & L6     & M6 V   & M6.5 V  & L5.5  & L3.5 \\

\hline
\multicolumn{6}{c}{Geballe et al. (2002)}\\
\hline

H$_2$O 1.5 $\mu$m  & L0     & $<$L0  & $<$L0   & $<$L0 & $<$L0 \\
CH$_4$ 2.2 $\mu$m  & L3     & L4     & $<$L3   & $<$L3 & L4 \\

\hline
\multicolumn{6}{c}{Burgasser et al. (2006,2007)}\\
\hline

H$_2$O-J           & L2.5   & L0      & L0.5   & L3.5  & L3.5 \\
H$_2$O-H           & M9.5   & $<$L0   & $<$L0  & L1    & L0 \\
CH$_4$-K           & L1.5   & L2.5    & L0.5   & $<$L0 & L3 \\

\hline
\multicolumn{6}{c}{Allers et al. (2007)}\\
\hline

H$_2$O             & L0.5   & M7.5 V  & M8.5 V & L3.5  & L2 \\

\hline
\multicolumn{6}{c}{Burgasser et al. (2008)}\\
\hline

H$_2$O($c$)        & L3.5   & M8 V    & L0     & L2.5  & L0 \\

\hline

Mean Sp. Type\tablenotemark{a}    & L1$\pm$1.5 & M8.5 V$\pm$2.5  & M9.5 V$\pm$1  & L3.5$\pm$1.5      &  L2.5$\pm$1.5 \\

\enddata
\tablenotetext{a}{Mean spectral type not including upper limits (e.g.,
  $<$L0).  The error is the RMS error.}

\end{deluxetable}

\clearpage

\begin{deluxetable}{llll}
  \tablecolumns{4}
  \tablenum{7}
\tabletypesize{\scriptsize} 
\tablewidth{0pc}
\tablecaption{\label{tab:SpTypes}Summary of Spectral Types for Objects Observed with SpeX}
\tablehead{
\colhead{} & 
\multicolumn{3}{c}{Spectral Types} \\
\cline{2-4}

\colhead{Object} &
\colhead{$i'-J$} &
\colhead{Direct Comparison} &
\colhead{Spectral Indices}}

\startdata
 
2MASS J1059+3042 & T8   & T4     & $\cdots$ \\
2MASS J1230+2827 & L0   & L2     & L1$\pm$1.5 \\
2MASS J1237+3028 & L0   & M7.5 V & M8.5$\pm$2.5 \\
2MASS J1253+2728 & M9 V & M8.5 V & M9.5 V$\pm$1 \\
2MASS J1431+1436 & L2   & L2     & L3.5$\pm$1.5 \\
2MASS J1434+2202 & L1   & sdM9   & L2.5$\pm$1.5 \\

\enddata

\end{deluxetable}

\clearpage

\newpage

\newpage

\newpage

\newpage

\newpage

\newpage

\newpage

\newpage

\begin{references}

\reference{Aba03} Abazajian, D., Adelman, J., Agueros, M. et al. 2003, AJ, 126, 2081

\reference{Aba03} Abazajian, D., Adelman, J., Agueros, M. et al. 2004, AJ, 128, 502

\reference{Aba03} Abazajian, D., Adelman, J., Agueros, M. et al. 2005, AJ, 129, 1755

\reference{Ade06} Adelman-McCarthy, J., Agueros, M., Allam, S. et al. 2006, ApJS, 162, 38

\reference{Ade07} Adelman-McCarthy, J., Agueros, M., Allam, S. et al. 2007, ApJS, 172, 634

\reference{2007ApJ...657..511A} Allers, K.~N., et al.\ 2007, \apj, 657, 511 

\reference{Art06} Artigau, E., Doyon, R., Lafreniere, D., Nadeau, D., Robert, J. and Albert, L. 2006, ApJ, 651, L57

\reference{Bur02*} Burgasser, A., Kirkpatrick, J., Brown, M., et al. 2002, ApJ, 564, 421

\reference{Bur03a} Burgasser, A., Kirkpatrick, J., McElwain, M., Cutri, R., Burgasser, A. and Skrutskie, M. 2003a, AJ, 125, 850

\reference{Bur03b*} Burgasser, A., McElwain, M., Kirkpatrick, J. 2003b, AJ, 126, 2487

\reference{Bur04*} Burgasser, A., McElwain, M., Kirkpatrick, J., Cruz, K., Tinney, C. and Reid, N. 2004a, AJ, 127, 2856

\reference{Bur04} Burgasser, A. et al. 2004b, ApJ, 604, 827

\reference{Bur06}  Burgasser, A., Geballe, T., Leggett, S., Kirkpatrick, J., Golimowski, D. 2006, ApJ, 637, 1067

\reference{Bur07} Burgasser, A., Cruz, K. and Kirkpatrick, J. 2007a, ApJ, 657, 494

\reference{2007ApJ...658..557B} Burgasser, A.~J., Looper, D.~L., Kirkpatrick, J.~D., \& Liu, M.~C.\ 2007b, \apj, 658, 557 


\reference{Bur08}  Burgasser, A., Looper, D., Kirkpatrick, J., Cruz, K. and Swift, B. 2008a, ApJ, 674, 451

\reference{Bur08b} Burgasser 2008b, submitted

\reference{2008ApJ...674..451B} Burgasser, A.~J., Looper, D.~L., Kirkpatrick, J.~D., Cruz, K.~L., \& Swift, B.~J.\ 2008, \apj, 674, 451 

\reference{Burr06} Burrows, A. et al. 2006, ApJ, 640, 1063

\reference{Chi05} Chiu, K., Fan, X., Leggett, S., Golimowski, D., Zheng, W., Geballe, T., Schneider, D. and Brinkmann, J. 2006, AJ, 131, 2722

\reference{Chi08} Chiu, K., Liu, M., Jiang, L. et al. 2008, MNRAS, 385, L53

\reference{Cru03} Cruz, K., Reid, N., Liebert, J., Kirkpatrick, J. and Lowrance, P. 2003, AJ, 126, 2421

\reference{Cru07} Cruz, K., Reid, N., Kirkpatrick, J., et al. 2007, AJ, 133, 439

\reference{2004PASP..116..362C} Cushing, M.~C., Vacca, W.~D., and Rayner, J.~T. 2004, PASP, 116, 362 

\reference{Cus05} Cushing, M., Rayner, J., and Vacca, W. 2005, ApJ, 623, 1115

\reference{Dah02} Dahn, C. et al. 2002, AJ, 124, 1170

\reference{Dea05} Deacon, N., Hambly, N. and Cooke, J. 2005, AA, 435, 363

\reference{Del08} Delorme, P. et al. 2008, AA, 484, 469

\reference{2000AJ....119..928F} Fan, X., et al.\ 2000, \aj, 119, 928 

\reference{Fin07} Finch, C., Henry, T., Subasavage, J., Jao, W. and Hambly, N. 2007, AJ, 134, 252

\reference{Geb02}  Geballe, T., Knapp, G., Leggett, S. et al. 2002, ApJ, 564, 466

\reference{Giz00}  Gizis, J., Monet, D., Reid, N. et al. 2000, AJ, 120, 1085

\reference{Gra03} Granvik, M., Virtanen, J., Muinonen, K., Bowell, E., Koehn, B. and Tancredi, G. 2003, Earth, Moon and Planets, 92, 73

\reference{Ham04} Hambly, N., Henry, T., Subasavage, J., Brown, M., and Jao, W. 2004, AJ, 128, 437

\reference{Har03} Harris, H., Conard, D., Frederick, V. et al. 2003, IAU Symposium 211, 409

\reference{Haw02} Hawley, S., Covey, K., Knapp, G. et al. 2002, AJ, 123, 3409

\reference{Hen97} Henry, T., Ianna, P., Kirkpatrick, J. and Jahreiss, H. 1997, ApJ, 114, 388

\reference{Hen02} Henry, T., Walkowicz, L., Barto, T. and Golimowski, D. 2002, AJ, 123, 2002

\reference{Jam08} Jameson, R., Casewell, S., Bannister, N. et al. 2008, MNRAS, 384, 1399

\reference{Ken04} Kendall, T., Delfosse, X., Martin, E. and Forveille, T. 2004, AA, 416, L17

\reference{Ken07} Kendall, T., Jones, H., Pinfield, D. et al. 2007, MNRAS, 374, 445

\reference{1991ApJS...77..417K} Kirkpatrick, J.~D., Henry, T.~J., \& McCarthy, D.~W., Jr.\ 1991, \apjs, 77, 417 

\reference{Kir99} Kirkpatrick, J., Reid, N., Liebert, J. et al. 1999, ApJ, 519, 802

\reference{Kir00} Kirkpatrick, J. et al. 2000, AJ, 120, 447

\reference{2008ASPC..384...85K} Kirkpatrick, J.~D.\ 2008, 
14th Cambridge Workshop on Cool Stars, Stellar Systems, and the Sun, 384, 
85 

\reference{Kna04} Knapp,G. et al. 2004, AJ, 127, 3553

\reference{Law07} Lawrence, A., Warren, S., Almaini, O. et al. 2007, MNRAS, 379, 1599

\reference{2002ApJ...564..452L} Leggett, S.~K., et al.\ 2002, \apj, 564, 452 

\reference{Lup03} Lepine, S., Shara, M. and Rich, R. 2003, AJ, 126, 921

\reference{Lep08} Lepine, S. 2008, AJ, 135, 2177

\reference{Lep08} Lepine, S. and Scholz, R. 2008, ApJ, 681, L33

\reference{Lie06} Liebert, J. and Gizis, J. 2006, PASP, 118, 659

\reference{Liu05} Liu, M. and Legget, S. 2005, ApJ, 634, 616

\reference{Lod05} Lodieu, N., Scholz, R., McCaughrean, M., Ibata, R,, Irwin, M. and Zinnecker, H. 2005, AA, 440, 1061

\reference{Loo07} Looper, D., Kirkpatrick, J. and Burgasser, A. 2007, AJ, 134, 1162

\reference{Loo08} Looper, D., et al. 2008, ApJ, in press (astro-ph 0806.1059)

\reference{Luy79} Luyten, W. 1979, LHS Catalogue (2nd ed.; Minneapolis: Univ. Minnesota Press)

\reference{Mar08} Marshall, J. 2008, AJ, 135, 1000

\reference{Met08} Metchev, S. Kirkpatrick, D., Berriman, B., Looper, D. 2008, ApJ, 676, 1281

\reference{Opp01} Oppenheimer, B., Hambly, N., Digby, A., Hodgkin, S. and Saumon, D. 2001, Science, 292, 698

\reference{Pha07} Phan-Bao, N., Bessell, M., Martin, E. et al. 2008, MNRAS, 383, 831

\reference{Pie03} Pier, J., Munn, J., Hindsley, R., Hennessy, G., Kent, S., Lupton, R. and Ivezic, Z. 2003, AJ, 125, 1559

\reference{2008arXiv0806.0294P} Pinfield, D.~J., et al.\ 2008, ArXiv e-prints, 806, arXiv:0806.0294 

\reference{Por04} Pokorny, R., Jones, H., Hambly, N. and Pinfield, D. 2004, AA, 421, 763

\reference{2003PASP..115..362R} Rayner, J.~T., Toomey, D.~W., Onaka, P.~M., Denault, A.~J., Stahlberger, W.~E., Vacca, W.~D., Cushing, M.~C., and Wang, S. 2003, PASP, 115, 362 

\reference{2001AJ....121.1710R} Reid, I.~N., Burgasser, A.~J., Cruz, K.~L., Kirkpatrick, J.~D., \& Gizis, J.~E.\ 2001, \aj, 121, 1710 

\reference{Rei04} Reid, N., Cruz, K., Allen, P. et al. 2004, AJ, 128, 463

\reference{Rei07} Reid, N., Cruz, K. and Allen, P. 2007, AJ, 133, 2825

\reference{Rui01} Ruiz, M., Wischnjewsky, M., Rojo, P. and Gonzalez, L. 2001, ApJS, 133, 119

\reference{Sal02} Salim, S. and Gould, A. 2002, ApJ, 575, 83

\reference{Sch07} Schmidt, S., Cruz, K., Bongiorno, B., Liebert, J. and Reid, N. 2007, 133, 2258

\reference{Sch03} Scholz, R., Lehmann, I., Matute, I. and Zinnecker, H. 2004, AA, 425, 519

\reference{Skr06} Skrutskie, M., Cutri, R., Stiening, R. et al. 2006, AJ, 131, 1163

\reference{Sub05} Subasavage, J., Henry, T., Hambly, N., Brown, M. and Wei-Chun, J. 2005, AJ, 129, 413

\reference{Tee03} Teegarden, B. et al. 2003, ApJ, 589, L51

\reference{2001ApJ...552L.147T} Testi, L., et al.\ 2001, \apjl, 552, L147 


\reference{Tin05} Tinney, C., Burgasser, A., Kirkpatrick, J. and McElwain, M. 2005, AJ, 130, 2326

\reference{Tsu96} Tsuji, T., Ohnaka, K. and Aoki, W. 1996, AA, 305, L1

\reference{2003PASP..115..389V} Vacca, W.~D., Cushing, M.~C., and Rayner, J.~T. 2003, PASP, 115, 389 

\reference{Vra04} Vrba, F., Henden, A., Luginbuhl, C. et al. 2004, AJ, 127, 2948

\reference{Wil03} Wilson, J., Miller, N., Gizis, J. et al. 2003, IAU Symposium 211, 197

\reference{Wro01} Wroblewski, H. and Costa, E. 2001, AA, 367, 725

\reference{Yor03} York, D., Adelman, J., Anderson, J. et al. 2000, AJ, 120, 1579

\end{references}
\end{document}